\documentclass[12pt]{article}
\usepackage{color}

\newcommand{\beq}{\begin{equation}}
\newcommand{\eeq}{\end{equation}}
\newcommand{\beqa}{\begin{eqnarray}}
\newcommand{\eeqa}{\end{eqnarray}}
\newcommand{\beqar}{\begin{eqnarray*}}
\newcommand{\eeqar}{\end{eqnarray*}}

\newcommand{\al}{\alpha}
\newcommand{\be}{\beta}

\def\spa          {\ \ \ }
\def\non          {\nonumber}
\def\ha           {\mbox{$\frac{1}{2}$}}

\def\s  {\sigma}
\def\spa          {\ \ \ }
\def\mand         {\spa\mbox{and}\spa}

\def\Tr           {\mbox{\rm Tr}\,}

\def\cd           {{\cdot}}
\def\ran          {\rangle}
\def\lan          {\langle}
\def\fsC    {C\!\!\!\!/\,}
\def\fsH    {H\!\!\!\!/\,}
\newcommand{\del}{\delta}

\newcommand{\eps}{\epsilon}
\newcommand{\ga}{\gamma}

\newcommand{\inn}{\!\cdot\!}

\newcommand{\lam}{\lambda}

\newcommand{\z}{\zeta}

\newcommand{\labell}[1]{\label{#1}} 
\newcommand{\reef}[1]{(\ref{#1})}
\newcommand\prt{\partial}
\newcommand\veps{\varepsilon}

\newcommand\cL{{\cal L}}
\newcommand\cD{{\cal D}}

\def\sst#1{{\scriptscriptstyle #1}}
\def\0{{\sst{(0)}}}
\def\1{{\sst{(1)}}}
\def\2{{\sst{(2)}}}
\def\3{{\sst{(3)}}}
\def\4{{\sst{(4)}}}
\def\5{{\sst{(5)}}}
\def\6{{\sst{(6)}}}
\def\7{{\sst{(7)}}}
\def\8{{\sst{(8)}}}

\begin{document}
\baselineskip 18pt%
\begin{titlepage}
\vspace*{1mm}%
\hfill
\vbox{

    \halign{#\hfil         \cr
           } 
      }  
\vspace*{8mm}
\vspace*{8mm}%

\center{ {\bf \Large On RR Couplings, Singularity Structures and all order $\alpha'$ contact interactions to BPS String Amplitudes

}}\vspace*{3mm} \centerline{{\Large {\bf  }}}
\vspace*{5mm}
\begin{center}
{Ehsan Hatefi$^{1,2,3}$ }

\vspace*{0.6cm}{\small $^{1}$  Centre for Research in String Theory, School of Physics and Astronomy,
Queen Mary University of London, Mile End Road, London E1 4NS, UK},
\vskip.06in

{\small $^{2}$ National Institute for Theoretical Physics ,
School of Physics  and Mandelstam Institute for Theoretical Physics,University of the Witwatersrand, Wits, 2050, SA},
\vskip.06in
{\small $^{3}$ Institute des Hautes Etudes Scientifiques Bures-sur-Yvette, F-91440, France }

\vspace*{.3cm}
\end{center}
\begin{center}{\bf Abstract}\end{center}
\begin{quote}

We evaluate five point world-sheet string theory amplitudes of one transverse scalar field, two world volume gauge fields ( and two transverse scalars, a gauge field)  in the presence of a closed string Ramond-Ramond vertex operator in its symmetric picture. We carry out all the entire S-matrix elements of five point mixed RR-scalars/gauge fields $<C^{-1}\phi ^{0}A^{-1} A^{0}>$, $<C^{-1}\phi ^{-1}A^{0} A^{0}>$, $<C^{-1}A^{0}\phi ^{-1}\phi^{0}>$ and $<C^{-1}A^{-1}\phi ^{0}\phi^{0}>$ in detail and start comparing all order $\alpha'$  contact interactions and singularities  in both transverse and world volume directions. We explore the presence of various new couplings in string theory effective actions and find out their all order $\alpha'$ higher derivative corrections in both type IIA and IIB. Ultimately we make various remarks  for the singularities and  contact terms whose RR momenta are embedded in transverse directions.  $\alpha'$ corrections to some of Myers terms are also addressed.

 \end{quote}
\end{titlepage}

\section{Introduction}

The fundamental objects in string theory or the so called D-branes have been playing a key ingredient in various research topics on theoretical high energy physics  as well as  in super string theory \cite{Polchinski:1995mt}\cite{Witten:1995im}\cite{Polchinski:1996na}.

\vskip.05in

Indeed either a BPS or non-BPS D$_p$-brane includes  $(p+1)$-dimensional world volume fields  which must be thought of a hypersurface like in a ten dimensional flat space time.  We need to take into account some special boundary conditions to them, namely either Neumann or Dirichlet, depending on whether we apply  those boundary conditions through transverse or its world volume fields \cite{Polchinski:1994fq}. Note that  recently some remarks for brane-anti brane have also been mentioned in \cite{Hatefi:2015gwa}.

\vskip.1in

To have more complete picture of the effective actions of string theory and what has been carried out up to now, we just  point out to various papers that are important to the author. Myers in \cite{Myers:1999ps} did explore the form of a bosonic action which holds for  multiple D$_p$-brane configurations and the generalization of Myers action with its all order $\alpha'$ corrections (using the mixed open-closed scattering amplitudes) has been done in \cite{Hatefi:2012zh}. Having performed \cite{Hatefi:2012zh}, some new couplings were obtained. These new couplings  are not inside Effective field theory (EFT) and their importance has  played the fundamental role not only in performing
the ADM  reduction of IIB and exploring  dS brane world-volume solutions \cite{Hatefi:2012bp} but also in deriving $N^3$ entropy of $M5$ branes.  These couplings  could have some specific role in super gravity solutions as well  where the particular emphasis is paid on the near-extremal black-branes to actually get  to  $n^3$ entropy growth analysis \cite{Hatefi:2012sy}.

\vskip.1in

A remarkable paper  \cite{Howe:2006rv} on supersymmetrized version of that action was given.  A part of the supersymmetric action is known, in fact it involves symmetric traces of the non-abelian fields  and what needs exploration is further terms which do not belong to the category that we are looking for in this paper. 
Whereas the effective action for a bosonic brane  given by \cite{Leigh:1989jq} and naturally its supersymmetric one was written down by  \cite{Cederwall:1996pv}. One could read off a review of all the DBI, Wess-Zumino and Chern-Simons action just for BPS branes from  \cite{Hatefi:2010ik}. On the other hand, to reveal more about three standard ways of effective field theory of the D-brane action (which contain Taylor expansion-Myers Terms and Pull back), and to learn more about all sorts of higher derivative corrections of non-BPS  and BPS branes , we advise the section five of \cite{Hatefi:2012wj}.

\vskip 0.1in

It is also important to have some tools to actually deal with the mixed open-closed higher point functions of string amplitudes, where
one can refer to some of the pioneer works on either effective actions  or scattering amplitudes that are involved with several D$_p$-brane configurations as well as their string applications \cite{Hashimoto:1996bf}.

\vspace{.3in}

The paper is constructed as follows. In the next section we just introduce  vertex operators  with all details and notations and then we try to work out Type II super string computations with all order $\alpha'$ D-brane S-matrix of a Ramond-Ramond (RR) in symmetric picture, a scalar field in zero picture  with two world volume gauge fields on different pictures where we try to address the entire S-matrix and explain the whole techniques that are involved in that particular amplitude. \footnote{
We may wonder whether it is possible to apply T-duality to $<V_{C}V_{A} V_AV_A>$ S-matrix of \cite{Hatefi:2010ik} to get  to 
$<V_{C}V_{\phi} V_AV_A>$ S-matrix. Indeed as it has been explored there are various terms in the S-matrix of $<V_{C}V_{\phi} V_AV_A>$, that  carry momentum of RR in transverse direction that cannot be obtained by T-duality transformation in flat ten dimensions of space-time.  In fact the appearance of RR makes things subtle or complicated as argued in
\cite{Hatefi:2012ve}  and \cite{Park:2008sg} accordingly.}

\vskip.2in

 Afterwards we start comparing  all the contact interactions and singularity structures of $<V_{C}V_{\phi} V_AV_A>$ S-matrix in two different pictures in the presence of a  symmetric RR vertex operator. Basically we compare both all order $\alpha'$ contact interactions and all the singularity structures of $<C^{-1}\phi ^{0}A^{-1} A^{0}>$ with $<C^{-1}\phi ^{-1}A^{0} A^{0}>$, where the superscripts refer to the chosen picture of each string operator.  Although we regenerate all  $t,s,u,(t+s+u)$- channel poles in effective field theory, we also find out some new contact interaction and singularities in the $<C^{-1}\phi ^{0}A^{-1} A^{0}>$ S-matrix and for the first time , we explore their all order $\alpha'$ couplings in effective field theory  as well.\footnote{
There is the
possibility that some  of the terms derived in different pictures of the vertex
operators, might be related via Bianchi identities of the bulk. This would imply that some of the
contact interactions might be redundant but  not all.  In some of the specific examples ,  some of the  assumed  contact terms
seems to be reproduced by a specific combination of pull-back and Taylor
expansion of the CS terms. One might use some of the new terms to eliminate either the pull-back or the Taylor expansion.
Nevertheless, we  believe that not all the new couplings are redundant.  }

 \vspace{.2in}

 It is also worth reporting some sort of new singularities and new sort of Myers terms that appear in this particular picture  of  $<C^{-1}\phi ^{0}A^{-1} A^{0}>$ S-matrix where those new terms are actually the terms that carry momentum of RR in transverse direction and do involve $p.\xi$ terms inside the S-matrix elements. 
 \vspace{.1in} 
  
 Note that  these   $p.\xi$ terms  are derived by direct analysis of $<C^{-1}\phi ^{0}A^{-1} A^{0}>$, due to  non zero correlation function of RR field by the first term of scalar field 's vertex operator in zero picture, that is , all $<e^{ip.x(z)} \partial_i x^i(x_1)> $ terms are indeed non-zero. Therefore since  scalar field 's polarization is in the bulk , one expects to be concerned about  
all  $p.\xi$ terms and $ p^{i},p^{j}$ terms  whose momenta of RR are carried in transverse directions.  It is worth pointing out the  following fact as follows. Recently, it is shown in \cite{Hatefi:2015okf} that , if one does not know all the Bianchi identities of RR in the bulk, then certainly there will be no chance  to explore all the bulk singularities of non-BPS branes.
\vspace{.2in}

 We perform full comparisons  at each order of $\alpha'$ for all contact interactions as well, and that leads to finding out new couplings that can be derived by just S-matrix analysis not by any other tools to our knowledge.

 \vspace{.2in}
 
 The profound relation of open-closed string plays the crucial role in matching out all the singularities of string theory with EFT, as it has been shown that all order $\alpha'$ higher derivative corrections to SYM couplings produce all massless poles at $(t+s+u)$-channel poles through an RR coupling with various BPS open strings. It has also been emphasised that, this phenomenon could have played the major role  for finding the universality conjecture on $\alpha'$ corrections of string theory \cite{Hatefi:2012rx}.

  \vspace{.2in}

We carry out the same analysis (this time for an RR, two scalars and a gauge field) in type IIA and IIB super string theory for both $<C^{-1}A^{0}\phi ^{-1}\phi^{0}>$ and $<C^{-1}A^{-1}\phi ^{0}\phi^{0}>$ S-matrices where  we seem to find out the same $ t,s,u,(t+s+u)$-singularity structures in the presence of an RR, even number of scalar fields. However, we claim that various new contact interactions  appear in the S-matrix by considering both scalar fields in zero picture. Indeed we derive these new couplings, show  that these couplings can just be discovered from  $<C^{-1}A^{-1}\phi ^{0}\phi^{0}>$ S-matrix and  explore their all order $\alpha'$ corrections in effective field theory side. Finally we conclude by mentioning various remarks about these S-matrices in the conclusion section.

\section{Type II Super string Computations with all order $\alpha'$ D-brane couplings }

In this section we would like to carry out the Conformal Field Theory (CFT) technique to be able to explore not only  all the singularities but also all the infinite contact interactions of the mixture of a closed string RR (in its symmetric picture) and various BPS open string fields. Indeed our calculation makes sense at the level of a world-sheet five point mixed closed-open string amplitude which must be done on the upper half-complex plane. We find the entire S-matrix elements  which hold on both world-volume and transverse component of D-branes.

One might be interested in seeing various efforts that have been performed on both BPS and non-BPS amplitudes
\cite{Kennedy:1999nn,Chandia:2003sh,Hatefi:2013yxa}.

In order to find out the effective action of string theory one needs to deal with  or calculate the scattering amplitudes and naturally
the first step to do so, is to  fix a particular  picture of the vertices. Namely,  the sum of the  superghost charges must have been (-2) for disk amplitudes. 

\vskip.1in

In our notations we use $\mu,\nu = 0, 1,..., 9$ for  the whole spacetime, while  $ a, b, c = 0, 1,..., p$ for world volume space
and $i,j = p + 1,...,9$ for transverse directions.  Here we would like to insist on the calculations in  the presence of  symmetric picture of RR but for the completeness we point out all the different vertex operators in various pictures as follows:
\beqa
V_{\phi}^{(0)}(x) &=& \xi_{i}\bigg(\partial
X^i(x)+\alpha' ik\cd\psi\psi^i(x)\bigg)e^{\alpha' ik\cd X(x)},
\nonumber\\
V_{\phi}^{(-1)}(y) &=&\xi.\psi(y) e^{-\phi(y)} e^{\alpha' ik\cd X(y)},
\nonumber\\
V_{A}^{(0)}(x) &=& \xi_{a}\bigg(\partial
X^a(x)+ \alpha'iq\cd\psi\psi^a(x)\bigg)e^{ \alpha' iq\cd X(x)},
\nonumber\\
V_{A}^{(-1)}(y) &=&\xi_a\psi^a(y) e^{-\phi(y)} e^{\alpha'iq\cd X(y)}
\nonumber\\
V_{C}^{(-\frac{1}{2},-\frac{1}{2})}(z,\bar{z})&=&(P_{-}\fsH_{(n)}M_p)^{\al\be}e^{-\phi(z)/2}
S_{\al}(z)e^{i\frac{\alpha'}{2}p\cd X(z)}e^{-\phi(\bar{z})/2} S_{\be}(\bar{z})
e^{i\frac{\alpha'}{2}p\cd D \cd X(\bar{z})},\nonumber\\
V_{C}^{(-\frac{3}{2},-\frac{1}{2})}(z,\bar{z})&=&(P_{-}\fsC_{(n-1)}M_p)^{\al\be}e^{-3\phi(z)/2}
S_{\al}(z)e^{i\frac{\alpha'}{2}p\cd X(z)}e^{-\phi(\bar{z})/2} S_{\be}(\bar{z})
e^{i\frac{\alpha'}{2}p\cd D \cd X(\bar{z})},
\label{BPS12}
\eeqa

To our knowledge the vertex of RR in asymmetric picture has been first shown by an interesting paper on open string theory \cite{Bianchi:1991eu} and then it was argued with some more details  in \cite{Liu:2001qa} where  the following kinematic relations are also considered

\beqa
  k^2=q^2=p^2=0 \quad q.\xi=0 ,
\nonumber\eeqa
We also apply
 Doubling trick to make use of holomorphic components of world sheet fields as well, that is, 
\begin{displaymath}
\tilde{X}^{\mu}(\bar{z}) \rightarrow D^{\mu}_{\nu}X^{\nu}(\bar{z}) \ ,
\spa
\tilde{\psi}^{\mu}(\bar{z}) \rightarrow
D^{\mu}_{\nu}\psi^{\nu}(\bar{z}) \ ,
\spa
\tilde{\phi}(\bar{z}) \rightarrow \phi(\bar{z})\,, \mand
\tilde{S}_{\al}(\bar{z}) \rightarrow M_{\al}{}^{\be}{S}_{\be}(\bar{z})
 ,
\end{displaymath}
where
\begin{displaymath}
D = \left( \begin{array}{cc}
-1_{9-p} & 0 \\
0 & 1_{p+1}
\end{array}
\right) \ ,\,\, \mand
M_p = \left\{\begin{array}{cc}\frac{\pm i}{(p+1)!}\ga^{a_{1}}\ga^{a_{2}}\ldots \ga^{a_{p+1}}
\eps_{a_{1}\ldots a_{p+1}}\,\,\,\,{\rm for\, p \,even}\\ \frac{\pm 1}{(p+1)!}\ga^{a_{1}}\ga^{a_{2}}\ldots \ga^{a_{p+1}}\ga_{11}
\eps_{a_{1}\ldots a_{p+1}} \,\,\,\,{\rm for\, p \,odd}\end{array}\right.
\end{displaymath}

\vskip.1in

 Although all the details of spinor part have been verified in \cite{Hatefi:2012wj}, we just clarify the definitions of projector and RR's field strength as follows

\beqa
(P_{-}\fsH_{(n)})^{\al\be} =
C^{\al\del}(P_{-}\fsH_{(n)})_{\del}{}^{\be} \quad\quad, P_{-} = \ha (1-\ga^{11})\eeqa
and 
\begin{displaymath}
\fsH_{(n)} = \frac{a
_n}{n!}H_{\mu_{1}\ldots\mu_{n}}\ga^{\mu_{1}}\ldots
\ga^{\mu_{n}}
\ ,
\non\end{displaymath}
 where for IIA and IIB  we use  $n=2,4,a_n=i$ and  $n=1,3,a_n=1$ appropriately. 
 
 Here we just work out with the holomorphic parts of correlations  but the interested reader can easily find out
all the tricks in the Appendix part of  \cite{Hatefi:2012wj} as well.

\vskip 0.2in

\subsection{ All order $\alpha'$ S-matrix element of   $<C^{-1}\phi ^{0}A^{-1} A^{0}>$ }

The complete form of the S-matrix element of a closed string RR (in its symmetric picture) $n$-form  field strength and a transverse scalar field in zero picture and two world volume gauge fields  $<C^{-1}\phi ^{0}A^{-1} A^{0}>$ can be found by  the following correlation functions

\begin{eqnarray}
{\cal A}^{<C^{-1}\phi ^{0}A^{-1} A^{0}>} & \sim & \int dx_{1}dx_{2}dx_{3}dzd\bar{z}\,
  \lan V_{\phi}^{(0)}{(x_{1})}
V_{A}^{(-1)}{(x_{2})}V_A^{(0)}{(x_{3})}
V_{RR}^{(-\frac{1}{2},-\frac{1}{2})}(z,\bar{z})\ran,\labell{sstring11}\eeqa

We just look for  a special ordering. Setting the Wick theorem, the  amplitude  is written down as follows
\beqa {\cal A}^{<C^{-1}\phi ^{0}A^{-1} A^{0}>}&\sim& \int
 dx_{1}dx_{2}dx_{3}dx_{4} dx_{5}\,
(P_{-}\fsH_{(n)}M_p)^{\al\be}\xi_{1i}\xi_{2a}\xi_{3b}x_{45}^{-1/4}(x_{24}x_{25})^{-1/2}\nonumber\\&&
\times(I_1+I_2+I_3+I_4)\Tr(\lam_1\lam_2\lam_3),\labell{12511}\eeqa where
$x_{ij}=x_i-x_j$, $x_{4}=z$, $x_{5}=\bar z$ and

\beqa
I_1&=&{<:\partial X^i(x_1)e^{\alpha' ik_1.X(x_1)}:e^{\alpha' ik_2.X(x_2)}
:\partial X^b(x_3)e^{\alpha' ik_3.X(x_3)}:e^{i\frac{\alpha'}{2}p.X(x_4)}:e^{i\frac{\alpha'}{2}p.D.X(x_5)}:>}
 \  \non \\&&\times{<:S_{\al}(x_4):S_{\be}(x_5):\psi^a(x_2):>},\nonumber\\
I_2&=&{<:\partial X^i(x_1)e^{\alpha' ik_1.X(x_1)}:e^{\alpha' ik_2.X(x_2)}
:e^{\alpha' ik_3.X(x_3)}:e^{i\frac{\alpha'}{2}p.X(x_4)}:e^{i\frac{\alpha'}{2}p.D.X(x_5)}:>}
 \  \non \\&&\times{<:S_{\al}(x_4):S_{\be}(x_5)::\psi^a(x_2):\alpha' ik_3.\psi\psi^{b}(x_3)>},\nonumber\\
 I_3&=&{<: e^{\alpha' ik_1.X(x_1)}:e^{\alpha' ik_2.X(x_2)}
:\partial X^b(x_3)e^{\alpha' ik_3.X(x_3)}:e^{i\frac{\alpha'}{2}p.X(x_4)}:e^{i\frac{\alpha'}{2}p.D.X(x_5)}:>}
 \  \non \\&&\times{<:S_{\al}(x_4):S_{\be}(x_5):\alpha' ik_1.\psi\psi^{i}(x_1):\psi^a(x_2):>},\nonumber\\
 I_4&=&{<: e^{\alpha' ik_1.X(x_1)}:e^{\alpha' ik_2.X(x_2)}
:e^{\alpha' ik_3.X(x_3)}:e^{i\frac{\alpha'}{2}p.X(x_4)}:e^{i\frac{\alpha'}{2}p.D.X(x_5)}:>}
 \  \non \\&&\times{<:S_{\al}(x_4):S_{\be}(x_5)
:\alpha' ik_{1}\cd\psi\psi^i(x_1)::\psi^a(x_2):\alpha' ik_{3}\cd\psi\psi^b(x_3):>}.
\label{i1234}
\eeqa

\vskip 0.1in

We actually use the standard  propagators , as follows
\begin{eqnarray}
\lan X^{\mu}(z)X^{\nu}(w)\ran & = & -\frac{\alpha'}{2}\eta^{\mu\nu}\log(z-w) \ , \non \\
\lan \psi^{\mu}(z)\psi^{\nu}(w) \ran & = & -\frac{\alpha'}{2}\eta^{\mu\nu}(z-w)^{-1} \ ,\non \\
\lan\phi(z)\phi(w)\ran & = & -\log(z-w) \ 
\labell{prop342}\end{eqnarray}

We also need to take into account the Wick's theorem to be able to investigate all the bosonic correlators. To see further details , the section 3  of  \cite{Hatefi:2015jpa} is strongly suggested.

 Let us just address the most complicated fermionic correlation function of two spin operators/ two different currents and a fermion field , where  all the possible contractions have to be considered.

 Once again we use $x_{4}=z$, $x_{5}=\bar z$ . Note that unlike the open string correlator where integration is on the real line  $x_{4},x_{5}$ are integrated on the upper half plane. It is only for the purposes of the Wick contractions that we can forget the complex conjugation of one variable to another, in order to simplify things.

\beqa
I_6^{bcaid}&=&<:S_{\al}(x_4):S_{\be}(x_5):\psi^d\psi^i(x_1)::\psi^a(x_2):\psi^c\psi^b(x_3)>\nonumber\\
&=&\bigg\{(\Gamma^{bcaid}C^{-1})_{{\alpha\beta}}+\alpha' r_1\frac{Re[x_{14}x_{25}]}{x_{12}x_{45}}+\alpha' r_2\frac{Re[x_{14}x_{35}]}{x_{13}x_{45}}+\alpha' r_3\frac{Re[x_{24}x_{35}]}{x_{23}x_{45}}\nonumber\\&&+\alpha'^2 r_4(\frac{Re[x_{14}x_{35}]}{x_{13}x_{45}})(\frac{Re[x_{24}x_{35}]}{x_{23}x_{45}})
\bigg\}2^{-5/2}x_{45}^{5/4}(x_{14}x_{15}x_{34}x_{35})^{-1}(x_{24}x_{25})^{-1/2},
\label{hh11}
\eeqa
so that
\beqa
r_1&=&\bigg(\eta^{da}(\Gamma^{bci}C^{-1})_{\alpha\beta}\bigg),\nonumber\\
r_2&=&\bigg(-\eta^{cd}(\Gamma^{bai}C^{-1})_{\alpha\beta}
+\eta^{db}(\Gamma^{cai}C^{-1})_{\alpha\beta}\bigg),\nonumber\\
r_3&=&\bigg(-\eta^{ac}(\Gamma^{bid}C^{-1})_{\alpha\beta}+\eta^{ab}(\Gamma^{cid}C^{-1})_{\alpha\beta}\bigg),\nonumber\\
r_4&=&\bigg((\eta^{cd}\eta^{ab}-\eta^{bd}\eta^{ac})(\gamma^{i}C^{-1})_{\alpha\beta}\bigg)
\eeqa
Replacing the above correlators and  performing some simple algebraic computations, one can further simplify the amplitude and write it down in a closed form as follows
\beqa
{\cal A}^{<C^{-1}\phi^{0} A^{-1}A^{0}>}&\!\!\!\!\sim\!\!\!\!\!&\int dx_{1}dx_{2} dx_{3}dx_{4}dx_{5}(P_{-}\fsH_{(n)}M_p)^{\al\be}I\xi_{1i}\xi_{2a}\xi_{3b}x_{45}^{-1/4}(x_{24}x_{25})^{-1/2}\nonumber\\&&\times
\bigg(I_7^a( a^i_1a^b_2)+a^i_1a^{ba}_3+a^b_2a^{ai}_4-\alpha'^2 k_{1d}k_{3c}I_6^{bcaid}\bigg)\Tr(\lam_1\lam_2\lam_3)
\labell{amp3q},\eeqa
where
\beqa
I&=&|x_{12}|^{\alpha'^2 k_1.k_2}|x_{13}|^{\alpha'^2 k_1.k_3}|x_{14}x_{15}|^{\frac{\alpha'^2}{2} k_1.p}|x_{23}|^{\alpha'^2 k_2.k_3}|
x_{24}x_{25}|^{\frac{\alpha'^2}{2} k_2.p}
|x_{34}x_{35}|^{\frac{\alpha'^2}{2} k_3.p}|x_{45}|^{\frac{\alpha'^2}{4}p.D.p},\nonumber\\
a^i_1&=&ip^i \bigg(\frac{x_{54}}{x_{14}x_{15}}\bigg),\nonumber\\
a^b_2&=&ik_1^{b}\bigg(\frac{x_{14}}{x_{13}x_{34}}+\frac{x_{15}}{x_{35}x_{13}}\bigg)
+ik_2^{b}\bigg(\frac{x_{24}}{x_{34}x_{23}}+\frac{x_{25}}{x_{35}x_{23}}\bigg),\nonumber\\
a^{ba}_3&=& \bigg\{(\Gamma^{bca}C^{-1})_{\alpha\beta}+(-\alpha'\eta^{ac}(\gamma^{b}C^{-1})_{\alpha\beta}+\alpha'\eta^{ab}(\gamma^{c}C^{-1})_{\alpha\beta})\frac{Re[x_{24}x_{35}]}{x_{23}x_{45}}\bigg\}\nonumber\\&&\times
\alpha' ik_{3c}2^{-3/2}x_{45}^{1/4}(x_{34}x_{35})^{-1}(x_{24}x_{25})^{-1/2}
\nonumber\\
a^{ai}_4&=&\alpha' ik_{1d}2^{-3/2}x_{45}^{1/4}(x_{24}x_{25})^{-1/2}(x_{14}x_{15})^{-1} \bigg\{(\Gamma^{aid}C^{-1})_{\alpha\beta}+\alpha'\eta^{ad}(\gamma^{i}C^{-1})_{\alpha\beta}\frac{Re[x_{14}x_{25}]}{x_{12}x_{45}}\bigg\}
,\nonumber\\
I_7^a&=&<:S_{\al}(x_4):S_{\be}(x_5):\psi^a(x_2):>=2^{-1/2}x_{45}^{-3/4}(x_{24}x_{25})^{-1/2}
(\gamma^{a}C^{-1})_{\alpha\beta}.\nonumber
\eeqa

\vskip 0.2in

Now one  could use  the   SL(2,R) invariance of the S-matrix and to remove  the $V_{CKG}$  we do gauge fixing over the position of open strings at  zero, one  and infinity.   By doing gauge fixing as $(x_1=0,x_2=1,x_3=\infty)$ , one needs to address the following integration on the upper half plane over the position of RR
\beqa
 \int d^2 \!z |1-z|^{a} |z|^{b} (z - \bar{z})^{c}
(z + \bar{z})^{d}\nonumber
\eeqa
where   $a,b,c$ are the combinations of the following  Mandelstam variables

\beqa
s&=&\frac{-\alpha'}{2}(k_1+k_3)^2,\quad t=\frac{-\alpha'}{2}(k_1+k_2)^2,\quad u=\frac{-\alpha'}{2}(k_2+k_3)^2
\nonumber\eeqa

and the results of the integrations for  $ d= 0,1$ and $ d=2$  were obtained accordingly in \cite{Fotopoulos:2001pt} and  \cite{Hatefi:2012wj} .

\vskip.1in

 Therefore, the final form of the S-matrix in this particular picture to all orders in $\alpha'$  is obtained as follows
\beqa {\cal A}^{<C^{-1}\phi^{0} A^{-1}A^{0}>}&=&{\cal A}_{1}+{\cal A}_{2}+{\cal A}_{3}+{\cal A}_{4}+{\cal A}_{5}+{\cal A}_{6}+{\cal A}_{7}\labell{711u}\eeqa

where
\beqa
{\cal A}_{1}&\!\!\!\sim\!\!\!&2^{-1/2}\xi_{1i}\xi_{2a}\xi_{3b}
\bigg[-k_{3c}k_{1d}\Tr(P_{-}\fsH_{(n)}M_p\Gamma^{bcaid})
+k_{3c}p^i\Tr(P_{-}\fsH_{(n)}M_p\Gamma^{bca})\bigg]
4(-t-s-u)L_1,
\nonumber\\
{\cal A}_{2}&\sim&2^{-1/2}\Tr(P_{-}\fsH_{(n)}M_p \Gamma^{aid})\xi_{1i}\xi_{2a}k_{1d}
\bigg\{-2k_1.\xi_3 (ut)+2k_2.\xi_3 (st)
\bigg\}L_2\nonumber\\
{\cal A}_{5}&\sim&2^{-1/2}\Tr(P_{-}\fsH_{(n)}M_p \gamma^{i})\xi_{1i}
\bigg\{\xi_{3}.\xi_{2}(2ts)+2k_1.\xi_3(2k_3.\xi_2)t-4u k_1.\xi_2(k_1.\xi_3)+4sk_2.\xi_3k_1.\xi_2\bigg\}L_1\nonumber\\
{\cal A}_{4}&\sim&-2^{-1/2}(st)L_2
\bigg\{  \xi_{3b}\xi_{1i}\xi_{2a}\Tr(P_{-}\fsH_{(n)}M_p \Gamma^{bai})u+2k_3.\xi_2 k_{1d}\xi_{1i}\xi_{3b}\Tr(P_{-}\fsH_{(n)}M_p \Gamma^{bid})\nonumber\\&&
-\Tr(P_{-}\fsH_{(n)}M_p \Gamma^{cid})k_{1d}k_{3c}\xi_{1i}(2\xi_2.\xi_3)
\bigg\}
\nonumber\\
{\cal A}_{3}&\sim&-2^{-1/2}\xi_{1i}k_{3c}
\bigg\{-2k_1.\xi_2 \xi_{3b}\Tr(P_{-}\fsH_{(n)}M_p \Gamma^{bci})(us)+2k_1.\xi_3 \xi_{2a}\Tr(P_{-}\fsH_{(n)}M_p \Gamma^{cai})(ut)
\bigg\}L_2\nonumber\\
{\cal A}_{6}&\sim&2^{-1/2}(st) p^i \xi_{1i}
\bigg\{ 2k_3.\xi_2 \Tr(P_{-}\fsH_{(n)}M_p \gamma^{b})\xi_{3b}-2\xi_3.\xi_2 \Tr(P_{-}\fsH_{(n)}M_p \gamma^{c})k_{3c}
\bigg\}L_2\nonumber\\
{\cal A}_{7}&\sim&2^{-1/2} p^i \xi_{1i}\Tr(P_{-}\fsH_{(n)}M_p \gamma^{a})\xi_{2a}
\bigg\{ 2k_1.\xi_3 (ut)-2\xi_3.k_2(st)\bigg\}L_2\labell{483}\eeqa
where the functions
 $L_1,L_2,$ are
\beqa
L_1&=&(2)^{-2(t+s+u)-1}\pi{\frac{\Gamma(-u+\frac{1}{2})
\Gamma(-s+\frac{1}{2})\Gamma(-t+\frac{1}{2})\Gamma(-t-s-u)}
{\Gamma(-u-t+1)\Gamma(-t-s+1)\Gamma(-s-u+1)}},\nonumber\\
L_2&=&(2)^{-2(t+s+u)}\pi{\frac{\Gamma(-u)
\Gamma(-s)\Gamma(-t)\Gamma(-t-s-u+\frac{1}{2})}
{\Gamma(-u-t+1)\Gamma(-t-s+1)\Gamma(-s-u+1)}},
\label{Ls2345}
\eeqa

\vskip.2in

As we have expected by interchanging
$\xi_{2a}\rightarrow k_{2a}$ and also $\xi_{3b}\rightarrow k_{3b}$, the whole S-matrix  vanishes, which means that the amplitude does satisfy all the associated Ward identities  and the amplitude is non zero for various $p,n$ cases. Notice also we are dealing with all massless BPS strings  so the expansion is low energy expansion. This S-matrix does have all $t,s,u$ and particularly $(t+s+u)$ channel poles and in particular it has some extra singularities that are precisely carrying momentum of  RR in the bulk direction where we will show that these terms cannot be derived in the other picture ($<C^{-1}\phi^{-1} A^{0}A^{0}>$) S-matrix and we argue about them in the next section.

More significantly, in $<C^{-1}\phi^{0} A^{-1}A^{0}>$ S-matrix we discover the new form of contact interactions to all orders in $\alpha'$ that cannot be found in the other picture.
For the precise definitions of the expansions and more kinematical definitions and identities, one needs to look at  \cite{Hatefi:2012ve,Hatefi:2010ik}.

\vskip 0.2in

Note that by sending $t,s,u \rightarrow 0$ , one finds the expansion of the functions $L_1,ut L_2$
as follows 
 
\beqa
L_1&=&-{2^{-1}\pi^{5/2}}\left( \sum_{n=0}^{\infty}c_n(s+t+u)^n\right.
\left.+\frac{\sum_{n,m=0}^{\infty}c_{n,m}[s^n t^m +s^m t^n]}{(t+s+u)}\right.\nonumber\\
&&\left.+\sum_{p,n,m=0}^{\infty}f_{p,n,m}(s+t+u)^p[(s+t)^{n}(st)^{m}]\right)\nonumber\\
ut L_2 &=&-\pi^{3/2}\sum_{n=-1}^{\infty}b_n \frac{1}{s}(u+t)^{n+1}+\sum_{p,n,m=0}^{\infty}e_{p,n,m}s^{p}(tu)^{n}(t+u)^m
\labell{highcaap}.
\eeqa

with  the following  coefficients 
\beqa
&&b_{-1}=1,\,b_0=0,\,b_1=\frac{1}{6}\pi^2,\,b_2=2\z(3),c_0=0,c_1=-\frac{\pi^2}{6},\nonumber\\
&&e_{2,0,0}=e_{0,1,0}=2\z(3),e_{1,0,0}=\frac{1}{6}\pi^2,e_{1,0,2}=\frac{19}{60}\pi^4,e_{1,0,1}=e_{0,0,2}=6\z(3),\nonumber\\
&&e_{0,0,1}=\frac{1}{3}\pi^2,e_{3,0,0}=\frac{19}{360}\pi^4,e_{0,0,3}=e_{2,0,1}=\frac{19}{90}\pi^4,e_{1,1,0}=e_{0,1,1}=\frac{1}{30}\pi^4,\labell{577}\\
&&c_2=-2\z(3),
\,c_{1,1}=\frac{\pi^2}{6},\,c_{0,0}=\frac{1}{2},c_{3,1}=c_{1,3}=\frac{2}{15}\pi^4,c_{2,2}=\frac{1}{5}\pi^4,\nonumber\\
&&c_{1,0}=c_{0,1}=0,
c_{3,0}=c_{0,3}=0\,
,\,c_{2,0}=c_{0,2}=\frac{\pi^2}{6},c_{1,2}=c_{2,1}=-4\z(3),\nonumber\\
&&f_{0,1,0}=\frac{\pi^2}{3},\,f_{0,2,0}=-f_{1,1,0}=-6\z(3),f_{0,0,1}=-2\z(3),c_{4,0}=c_{0,4}=\frac{1}{15}\pi^4.\, \nonumber
\eeqa

\vskip 0.2in

Meanwhile  the result of the S-matrix in different picture of scalar field, that is, 

$<C^{-1}\phi^{-1} A^{0}A^{0}>$ S-matrix was derived in  \cite{Hatefi:2012ve}  to be as follows

\begin{eqnarray}
{\cal A}^{<C^{-1}\phi^{-1} A^{0}A^{0}>} & \sim & \int dx_{1}dx_{2}dx_{3}dzd\bar{z}\,
  \lan V_{\phi}^{(-1)}{(x_{1})}
V_{A}^{(0)}{(x_{2})}V_A^{(0)}{(x_{3})}
V_{RR}^{(-\frac{1}{2},-\frac{1}{2})}(z,\bar{z})\ran,\nonumber\eeqa

\beqa {\cal A}^{<C^{-1}\phi^{-1} A^{0} A^{0}>}&=&{\cal A}_{1}+{\cal A}_{2}+{\cal A}_{3}+{\cal A}_{4}+{\cal A}_{5}\labell{711u}\eeqa
where
\beqa
{\cal A}_{1}&\!\!\!\sim\!\!\!&-2^{-1/2}\xi_{1i}\xi_{2a}\xi_{3b}
\bigg[k_{3d}k_{2c}\Tr(P_{-}\fsH_{(n)}M_p\Gamma^{bdaci})
\bigg]
4(-t-s-u)L_1,
\nonumber\\
{\cal A}_{2}&\sim&2^{-1/2}\Tr(P_{-}\fsH_{(n)}M_p \Gamma^{bdi})\xi_{1i}\xi_{3b}k_{3d}
\bigg\{2k_1.\xi_2 (us)-2k_3.\xi_2 (st)
\bigg\}L_2\nonumber\\
{\cal A}_{3}&\sim&-2^{-1/2}\Tr(P_{-}\fsH_{(n)}M_p \Gamma^{aci})\xi_{1i}\xi_{2a}k_{2c}
\bigg\{-2k_2.\xi_3 (st)+2k_1.\xi_3 (ut)
\bigg\}L_2\\
{\cal A}_{4}&\sim&-2^{-1/2}(st)L_2
\bigg\{ \xi_{3b}\xi_{1i}\xi_{2a}\Tr(P_{-}\fsH_{(n)}M_p \Gamma^{bai})u +2k_2.\xi_3 k_{3d}\xi_{1i}\xi_{2a}\Tr(P_{-}\fsH_{(n)}M_p \Gamma^{dai})
\nonumber\\&&+2k_3.\xi_2 k_{2c}\xi_{1i}\xi_{3b}\Tr(P_{-}\fsH_{(n)}M_p \Gamma^{bci})-\Tr(P_{-}\fsH_{(n)}M_p \Gamma^{dci})k_{3d}k_{2c}\xi_{1i}(2\xi_2.\xi_3)
\bigg\}\nonumber\\
{\cal A}_{5}&\sim&2^{-1/2}\Tr(P_{-}\fsH_{(n)}M_p \gamma^{i})\xi_{1i}
\bigg\{
\xi_{3}.\xi_{2}(2ts)+2k_1.\xi_3(2k_3.\xi_2)t-4u k_1.\xi_2(k_1.\xi_3)+4sk_2.\xi_3k_1.\xi_2\bigg\}L_1\nonumber
\labell{483}\eeqa


Let us first compare the results of the same S-matrix in different pictures and then start producing all the singularity structures as well as new contact interactions.

\section{Comparison on Singularity Structures of $<C^{-1}\phi^{0}A^{-1}A^{0}>$ with $<C^{-1}\phi^{-1}A^{0} A^{0}>$ }

First of all note that both ${\cal A}_{5}$'s in two different pictures are exactly matched.
The first term  ${\cal A}_{3}$ of $<C^{-1}\phi^{0}A^{-1}A^{0}>$ is exactly the first term  ${\cal A}_{2}$ of $<C^{-1}\phi^{-1}A^{0} A^{0}>$. Now if we add the 2nd term   ${\cal A}_{2}$ of  $<C^{-1}\phi^{-1}A^{0} A^{0}>$ with the 3rd term  
${\cal A}_{4}$ of  $<C^{-1}\phi^{-1}A^{0} A^{0}>$ and apply momentum conservation along the world volume of brane we  get

\beqa
-2^{-1/2} st L_2 (2k_3.\xi_2)\xi_{1i}\xi_{3b}\Tr(P_{-}\fsH_{(n)}M_p\Gamma^{bci}) (-k_{1c}-p_{c})\nonumber\eeqa

Now if we use the identity that has been found in  \cite{Hatefi:2015gwa}, that is,

\beqa  
p_c \epsilon^{a_{0}...a_{p-2}bc}=0 \label{BI1}\eeqa
 then we get to know the fact that  the first term of above equation precisely produces the 2nd term ${\cal A}_{4}$ of  $<C^{-1}\phi^{0}A^{-1} A^{0}>$. 
 
 \vskip.1in

 One can see the derivation of \reef{BI1} in various equations of \cite{Hatefi:2015gwa}. For example, it is shown in equation (9) of 
 \cite{Hatefi:2015gwa} that,  in order to get to the same result of three point function of one RR and a scalar field in both $<C^{-1}\phi^{-1}>$ and  $<C^{-2}\phi^{0}>$ S-matrix , the  equation (9) of \cite{Hatefi:2015gwa} or \reef {BI1} must hold. Another example to prove that \reef {BI1} holds is as follows. 
 It is shown in section five of   \cite{Hatefi:2015gwa} that, to get to the same result of four point function of  $<C^{-1}T^{0}\phi^{-1}>$ and  $<C^{-2}T^{0}\phi^{0}>$ S-matrix , the  equation \reef {BI1} must hold ( see the footnote of 19 in page 14 of \cite{Hatefi:2015gwa}).
 It has also been discovered  that   the amplitude of $<C^{-1}A^{-1}T^{0} T^{0}>$ satisfies Ward identity associated to the gauge field if and only if the above identity \reef{BI1} holds.

\vskip.1in

Likewise if we add the 1st term  ${\cal A}_{3}$  with the 2nd term  ${\cal A}_{4}$ of $<C^{-1}\phi^{-1}A^{0} A^{0}>$ and also apply momentum conservation we find the following elements 
\beqa
-2^{-1/2} st L_2 (2k_2.\xi_3)\xi_{1i}\xi_{2a}\Tr(P_{-}\fsH_{(n)}M_p\Gamma^{aci}) (k_{1c}+p_{c})\nonumber\eeqa
Once more one needs to apply the equation  $p_c \epsilon^{a_{0}...a_{p-2}ac}=0$ so that the first term of above precisely produces the 2nd term  ${\cal A}_{2}$ of  $<C^{-1}\phi^{0}A^{-1} A^{0}>$.

\vskip.1in

Simultaneously if we add the first term  ${\cal A}_{2}$ with the 2nd term  ${\cal A}_{3}$ of $<C^{-1}\phi^{0}A^{-1} A^{0}>$ with keeping in mind momentum conservation and  $p_c \epsilon^{a_{0}...a_{p-2}ac}=0$, we  then precisely produce the 2nd term  ${\cal A}_{3}$ of  $<C^{-1}\phi^{-1}A^{0} A^{0}>$.

\vskip.2in

Finally the last term  ${\cal A}_{4}$ of  $<C^{-1}\phi^{-1}A^{0} A^{0}>$ is exactly equivalent with the last term  ${\cal A}_{4}$ of  $<C^{-1}\phi^{0}A^{-1} A^{0}>$.\footnote{ Notice to momentum conservation and $p_c \epsilon^{a_{0}...a_{p-2}ac}=0$.}

\vskip.2in

 Therefore the upshot is that we can precisely produce all the singularities of $<C^{-1}\phi^{-1}A^{0} A^{0}>$ by $<C^{-1}\phi^{0}A^{-1} A^{0}>$ S-matrix (as we will show later on), however,  we have also some extra  contact interactions and other singularities  of $<C^{-1}\phi^{0}A^{-1} A^{0}>$ S-matrix (in the zero picture of  scalar field in the presence of a symmetric RR) that are absent in $<C^{-1}\phi^{-1}A^{0} A^{0}>$ S-matrix and we will argue about them in a moment.
 
\vskip.1in

In fact from the direct calculations we observe the facts  that at pole levels the whole ${\cal A}_{6}$ and ${\cal A}_{7}$ of   $<C^{-1}\phi^{0}A^{-1} A^{0}>$ S-matrix are extra terms that cannot be derived from direct computations of $<C^{-1}\phi^{-1}A^{0} A^{0}>$ S-matrix. 
\vskip.1in

Moreover, the  2nd contact interaction  ${\cal A}_{1}$ of   $<C^{-1}\phi^{0}A^{-1} A^{0}>$ is also extra term  that cannot be  derived from direct computations of $<C^{-1}\phi^{-1}A^{0} A^{0}>$ S-matrix on upper half plane, as we further elaborate on this coupling in the other section.  Let us first produce the different singularity structures.

\vskip.2in

We  do have massless scalar poles in t,s  and $(t+s+u)$ channels  as well as u channel gauge field poles. Here  we just produce the s-channel scalar poles and finally by interchanging  $2 \leftrightarrow 3$ and exchanging the respected momenta and polarisations we can produce t- channel poles as well.
If we replace the desired expansion of $ut L_2$ , we then obtain all  s-channel poles of string amplitude as follows (normalization constant is $(2\pi)^{1/2} m_p$)

\beqa
\frac{(2\pi\alpha')^2}{ p!}\mu_p \xi_{1i}\xi_{2a}k_{2c}\eps^{a_{0}\cdots a_{p-2}ac}H^{i}_{a_{0}\cdots a_{p-2}}\sum_{n=-1}^{\infty}\frac{1}{s}{b_n(u+t)^{n+1}}(2k_1.\xi_3)
\Tr(\lam_1\lam_2\lam_3)\label{UI}
\eeqa
In order to produce all these massless s-channel scalars , one has to consider a field theory sub amplitude as  
\beqa
{\cal A}&=&V^i_{\alpha}(C_{p-1},A_2,\phi)G^{ij}_{\alpha\beta}(\phi)V^j_{\beta}(\phi,A_3,\phi_1)\label{vvcx33}
\eeqa 
where  by taking into account the kinetic term of scalar fields $ \frac{(2\pi\alpha')^2 }{2}D^a\phi^i D_a\phi_i$  one obtains the following vertex as well as scalar propagator 

\beqa
V^j_{\beta}(\phi,A_3,\phi_1)&=&-2ik_1.\xi_3(2\pi\alpha')^2 T_p  \xi_1^j\Tr(\lam_3\lam_1\lam_\beta)
\labell{fvertex78}\\
({G}^{\phi})^{ij}_{\alpha\beta}&=&-\frac{i\delta^{ij}\delta^{\alpha\beta}}{(2\pi\alpha')^2T_p s}
\nonumber
\eeqa

Now we need to consider the mixed Chern-Simons coupling and the so called Taylor expended of scalar field as follows
\beqa
S_{1}&=&{i}(2\pi\alpha')^2\mu_p\int d^{p+1}\s {1\over (p-1)!}(\veps^v)^{a_0\cdots a_{p}}
 \Tr\left(F_{a_{0}a_{1}}\phi^i\right)
\prt_iC^{(p-1)}_{a_2\cdots a_{p}}
\labell{interac3}
\eeqa
to actually derive the following vertex operator of an RR, an on-shell gauge field and an off-shell scalar field as 
\beqa
V^i_{\alpha}(C_{p-1},A_2,\phi)&=&\frac{i (2\pi\alpha')^2\mu_p}{(p)!}(\veps^v)^{a_0\cdots a_p}H^i{}_{a_2\cdots a_p}\xi_{2a_{1}}k_{2a_{0}}\Tr(\lam_2\lambda_\alpha)\label{vvcx22}\eeqa
where $V^j_{\beta}(\phi,A_3,\phi_1)$ is derived from the kinetic term of the scalar field and in particular it has no correction , hence to be able to produce all s-channel poles we need to propose the  higher derivative corrections to \reef{interac3} as follows

\beqa
S_{2}&=&\sum_{n=-1}^{\infty}b_n (\alpha')^{(n+1)}\mu_p\int d^{p+1}\s {1\over (p-1)!}(\veps^v)^{a_0\cdots a_{p}}
\nonumber\\&&\times\partial_{i}C_{p-1}\wedge D_{a_{1}}...D_{a_{n+1}} F D^{a_{1}}...D^{a_{n+1}}\phi^i
\label{vvcx}
\eeqa

Having taken \reef{vvcx}, we were able to derive  all order vertex operator of \reef{vvcx22} as

\beqa
V^i_{\alpha}(C_{p-1},A_2,\phi)&=&\frac{i (2\pi\alpha')^2\mu_p}{(p)!}(\veps^v)^{a_0\cdots a_p}H^i{}_{a_2\cdots a_p}\xi_{2a_{1}}k_{2a_{0}}\Tr(\lam_2\lambda_\alpha)\sum_{n=-1}^{\infty}b_n(\alpha'k_2.k)^{n+1}
\label{imb1}\eeqa

Now if we replace \reef{imb1} and \reef{fvertex78} inside  \reef{vvcx33} , then one is able to precisely regenerate all order s-channel singularities 
of string amplitude \reef{UI} in the effective field theory as well.

All the  u-channel gauge field poles of the string amplitude  are given as 
\beqa
 \mu_p(2\pi\alpha')^{2}\frac{1}{(p)!u}\eps^{a_{0}\cdots a_{p-2}bd}\xi_{1i}k_{1d}
H^i{}_{a_0\cdots a_{p-2}}\sum_{n=-1}^{\infty}b_n (s+t)^{n+1}(2k_3.\xi_2\xi_{3b}-2k_2.\xi_3\xi_{2b} +2\xi_3.\xi_2 k_{2b})\labell{amp0012}\eeqa

Note that these u-channel gauge field poles can be reconstructed in the effective field theory by the following field theory sub amplitude

\beqa
{\cal A}&=&V^a_{\alpha}(C_{p-1},\phi_1,A)G^{ab}_{\alpha\beta}(A)V^b_{\beta}(A,A_2,A_3),\label{amp00}
\eeqa
where the vertices are 
\beqa
V^a_{\alpha}(C_{p-1},\phi_1,A)&=&\frac{i(2\pi\alpha')^2\mu_p}{(p)!}(\veps^v)^{a_0\cdots a_{p-1}a}H^i{}_{a_1\cdots a_{p-1}}\xi_{1i}k_{a_{0}}\Tr(\lam_1\lambda_\alpha)\sum_{n=-1}^{\infty}b_n(t+s)^{n+1},\nonumber\\
V^b_{\beta}(A,A_2,A_3)&=&-iT_p(2\pi\alpha')^{2}\Tr(\lam_2\lam_3\lambda_\beta)\bigg[2k_2.\xi_3\xi_2^b
-2 k_3.\xi_2\xi_3^b+\xi_3.\xi_2(k_3-k_2)^b\bigg],\nonumber\\
G_{\alpha\beta}^{ab}(A)&=&\frac{i\delta_{\alpha\beta}\delta^{ab}}{(2\pi\alpha')^2 T_p
u},,\label{amp001}
\eeqa

where  all order corrections to  $V^a_{\alpha}(C_{p-1},\phi_1,A)$  have been derived from \reef{vvcx}. By replacing these vertices into the field theory amplitude \reef{amp00}, one exactly produces all  u-channel gauge field poles that appeared in \reef{amp0012}.
\vskip.2in

For the completeness we just produce all the $(s+t+u)$- channel singularities of the S-matrix in the field theory as well. To do so, first we replace the part of the expansion of $L_1$ (which has poles) inside ${\cal A}_{5}$ so that one gets all the poles in string amplitude as 
\beqa
&&8\pi^3\mu_p\frac{\eps^{a_{0}\cdots a_{p}}\xi_{1i}
H^{i}_{a_0\cdots a_{p}}}{(p+1)!(s+t+u)}\Tr(\lam_1\lam_2\lam_3)
\sum_{n,m=0}^{\infty}c_{n,m}(s^{m}t^{n}+s^{n}t^{m})\nonumber\\&&
\bigg[2st\xi_{2}.\xi_3+4t k_1.\xi_3 k_3.\xi_2+4s k_1.\xi_2 k_2.\xi_3-4u k_1.\xi_2 k_1.\xi_3\bigg]
\label{amphigh87}\eeqa

In order to produce these poles , one has to consider the following sub amplitude in field theory side 
\beqa
V_{\alpha}^{i}(C_{p+1},\phi)G_{\alpha\beta}^{ij}(\phi)V_{\beta}^{j}(\phi,\phi_1,
A_2,A_3)\label{vienna1}\eeqa
where the scalar propagator can be  found  by taking the kinetic term of scalar fields ($\frac{(2\pi\alpha')^2} {2}D^a\phi^iD_a\phi_i$) and the vertex of $V_{\alpha}^{i}(C_{p+1},\phi)$ is obtained by taking the following effective action through Taylor expansion of scalar field 
\beqa
(2\pi\alpha')i\mu_p\int d^{p+1}\sigma {1\over (p+1)!}
(\veps^v)^{a_0\cdots a_{p}}\,\Tr\left(\phi^i\right)\,
\prt_iC^{(p+1)}_{a_0\cdots a_{p}} \nonumber\eeqa
so that 

\beqa
G_{\alpha\beta}^{ij}(\phi) &=&\frac{-i\delta_{\alpha\beta}\delta^{ij}}{T_p(2\pi\alpha')^2
k^2}=\frac{-i\delta_{\alpha\beta}\delta^{ij}}{T_p(2\pi\alpha')^2
(t+s+u)},\nonumber\\
V_{\alpha}^{i}(C_{p+1},\phi)&=&i(2\pi\alpha')\mu_p\frac{1}{(p+1)!}(\veps^v)^{a_0\cdots a_{p}}
 H^{i}_{a_0\cdots a_{p}}\Tr(\lambda_{\alpha}).
\labell{Fey}
\eeqa

To be able to produce all  scalar massless poles of the string amplitude  to all
orders in $\alpha'$, one needs to know  all order vertex operator of two scalar two gauge field couplings $V_{\beta}^{j}(\phi,\phi_1,
A_2,A_3)$. This vertex operator can be found by employing  all order $\alpha'$ SYM couplings  \cite{Hatefi:2012ve} as follows
\beqa
(2\pi\alpha')^4\frac{1}{ 2 \pi^2}T_p\left(\alpha'\right)^{n+m}\sum_{m,n=0}^{\infty}(\cL_{1}^{nm}+\cL_{2}^{nm}+\cL_{3}^{nm}),\labell{highder}\eeqa
\beqa
&&\cL_{1}^{nm}=-
\Tr\left(\frac{}{}a_{n,m}\cD_{nm}[D_a \phi^i D^b \phi_i F^{ac}F_{bc}]+ b_{n,m}\cD'_{nm}[D_a \phi^i F^{ac} D^b \phi_i F_{bc}]+h.c.\frac{}{}\right),\nonumber\\
&&\cL_{2}^{nm}=-\Tr\left(\frac{}{}a_{n,m}\cD_{nm}[D_a \phi^i D^b \phi_i F_{bc}F^{ac}]+\frac{}{}b_{n,m}\cD'_{nm}[D_a \phi^i F_{bc} D^b \phi_i F^{ac}]+h.c.\frac{}{}\right),\nonumber\\
&&\cL_{3}^{nm}=\frac{1}{2}\Tr\left(\frac{}{}a_{n,m}\cD_{nm}[D_a \phi^i D^a \phi_i F^{bc}F_{bc}]+\frac{}{}b_{n,m}\cD'_{nm}[D_a \phi^i F_{bc} D^a \phi_i F^{bc}]+h.c\frac{}{}\right),\nonumber\eeqa
where the following definitions for all higher derivative operators have been considered
 \cite{Hatefi:2010ik}
\beqa
\cD_{nm}(EFGH)&\equiv&D_{b_1}\cdots D_{b_m}D_{a_1}\cdots D_{a_n}E  F D^{a_1}\cdots D^{a_n}GD^{b_1}\cdots D^{b_m}H,\nonumber\\
\cD'_{nm}(EFGH)&\equiv&D_{b_1}\cdots D_{b_m}D_{a_1}\cdots D_{a_n}E   D^{a_1}\cdots D^{a_n}F G D^{b_1}\cdots D^{b_m}H.\nonumber
\eeqa

Since the off-shell scalar field is abelin, one needs to consider just two permutations of $\Tr(\lam_1\lambda_{\beta}\lam_2\lam_3),
\Tr(\lambda_{\beta}\lam_1\lam_2\lam_3)$ to be able to derive all order vertex of  $ V_{\beta}^{j}(\phi,\phi_1,
A_2,A_3)$  from the above corrections \reef{highder} as below

\beqa
V_{\beta}^{j}(\phi,\phi_1,
A_2,A_3)&=&\xi_{1}^j\frac{I_8}{2\pi^2}(\alpha')^{n+m}(a_{n,m}+b_{n,m})
\bigg(\frac{}{}(k_3\inn k_1)^m(k_1\inn k_2)^n+(k_3\inn k)^m(k_2\inn k)^n
\nonumber\\&&+(k_1\inn k_3)^n(k_1\inn k_2)^m+(k\inn k_3)^n (k\inn k_2)^m
\bigg),\labell{verppaa}\eeqa

 with the following definition for $I_8$
 \beqa
I_8&=&(2\pi\alpha')^4T_{p}\Tr(\lam_1\lam_2\lam_3\lambda_{\beta})\bigg[\frac{st}{2}\xi_{2}.\xi_3+t k_1.\xi_3 k_3.\xi_2+s k_1.\xi_2 k_2.\xi_3-u k_1.\xi_2 k_1.\xi_3\bigg]\nonumber\eeqa

 where  $k$ is  off-shell scalar field's momentum, and some of the coefficients $a_{n,m}$
and $b_{n,m}$ ( $b_{n,m}$ is symmetric \cite{Hatefi:2010ik} ) are
\beqa
&&a_{0,0}=-\frac{\pi^2}{6},\,b_{0,0}=-\frac{\pi^2}{12},a_{1,0}=2\z(3),\,a_{0,1}=0,\,b_{0,1}=-\z(3),a_{1,1}=a_{0,2}=-7\pi^4/90,\nonumber\\
&&a_{2,2}=(-83\pi^6-7560\z(3)^2)/945,b_{2,2}=-(23\pi^6-15120\z(3)^2)/1890,a_{1,3}=-62\pi^6/945,\nonumber\\
&&\,a_{2,0}=-4\pi^4/90,\,b_{1,1}=-\pi^4/180,\,b_{0,2}=-\pi^4/45,a_{0,4}=-31\pi^6/945,a_{4,0}=-16\pi^6/945,\nonumber\\
&&a_{1,2}=a_{2,1}=8\z(5)+4\pi^2\z(3)/3,\,a_{0,3}=0,\,a_{3,0}=8\z(5),b_{1,3}=-(12\pi^6-7560\z(3)^2)/1890,\nonumber\\
&&a_{3,1}=(-52\pi^6-7560\z(3)^2)/945, b_{0,3}=-4\z(5),\,b_{1,2}=-8\z(5)+2\pi^2\z(3)/3,\nonumber\\
&&b_{0,4}=-16\pi^6/1890.\eeqa
They are computed in \cite{Hatefi:2010ik}.
Now if we use   momentum conservation , we get $k_3\inn k=k_2.k_1-(k^2)/2$ and $k_2\inn k=k_1.k_3-(k^2)/2$,
whereas  $k^2$ in \reef{verppaa} is cancelled with the $k^2$ in the denominator of the  propagator. Since  we just want to produce singularities , we are ignoring those contact terms and considering  \reef{verppaa} and \reef{Fey} inside \reef{vienna1} , one explores the sub amplitude in field theory as follows 
\beqa
&&16\pi\mu_p\frac{\eps^{a_{0}\cdots a_{p}}\xi_{1i}
H^{i}_{a_0\cdots a_{p}}}{(p+1)!(s+t+u)}\Tr(\lam_1\lam_2\lam_3)
\sum_{n,m=0}^{\infty}\bigg((a_{n,m}+b_{n,m})[s^{m}t^{n}+s^{n}t^{m}]\nonumber\\&&
\bigg[2st\xi_{2}.\xi_3+4t k_1.\xi_3 k_3.\xi_2+4s k_1.\xi_2 k_2.\xi_3-4u k_1.\xi_2 k_1.\xi_3\bigg]
\label{amphigh8}\eeqa

Now we show that all the poles in field theory of \reef{amphigh8} can be matched with string amplitude poles that appeared in \reef{amphigh87}. After omitting the common factors of both string and field theory we compare string amplitude with sub amplitude in field theory for various cases of 
 $n,m$.  For $n=m=0$, the amplitude \reef{amphigh8} does carry  the following factor
\beqa
-4(a_{0,0}+b_{0,0})&=&-4(\frac{-\pi^2}{6}+\frac{-\pi^2}{12})=\pi^2\nonumber\eeqa
where the corresponding term for the string amplitude carries  $(2\pi^2 c_{0,0})$ which is exactly equivalent to the factor of $\pi^2$ in field theory sub amplitude.  At $\alpha'$ order,  \reef{amphigh8} carries the following coefficient 
\beqa
-(a_{1,0}+a_{0,1}+b_{1,0}+b_{0,1})(s+t)&=&0\nonumber\eeqa
where the corresponding term for the string amplitude is now proportional to   $\pi^2(c_{1,0}+c_{0,1})(s+t)$ which is zero as appeared in the field theory sub amplitude.  At $(\alpha')^2$ order, \reef{amphigh8} has the following numerical factor
\beqa
&&-4(a_{1,1}+b_{1,1})st-2(a_{0,2}+a_{2,0}+b_{0,2}+b_{2,0})[s^2+t^2]\nonumber\\
&&=\frac{\pi^4}{3}(st)+\frac{\pi^4}{3}(s^2+t^2)
\nonumber\eeqa

where the corresponding term for the string amplitude is now proportional to  $\pi^2 [c_{1,1}(2st)+(c_{2,0}+c_{0,2})(s^2+t^2)]$,  which is exactly equivalent to the factor of  field theory sub amplitude. The comparisons at orders of  $(\alpha')^3,(\alpha')^4$ are also done in \cite{Hatefi:2012ve}. Hence, one can keep comparing to all orders and show that indeed  all singularities of $(t+s+u)$ channels of string amplitude can be precisely reconstructed by the above field theory sub amplitudes.

\vskip.1in

Before further analysis let us compare all order contact interactions on two different pictures , start finding new coupling in the string theory effective action and also explore its all order $\alpha'$ corrections.

\section{Comparison on  Contact interactions to all $\alpha'$ orders   }

 If we look at the whole S-matrix elements in two different pictures  and  apply momentum conservation to the
 1st term  ${\cal A}_{1}$ of $<C^{-1}\phi^{0}A^{-1} A^{0}>$ we obtain the following elements

\beqa
-2^{3/2}(-t-s-u)L_1 \xi_{1i}\xi_{2a}\xi_{3b}
k_{3c}(-k_{3d}-k_{2d}-p_{d})\Tr(P_{-}\fsH_{(n)}M_p\Gamma^{bcaid})\nonumber\\=2^{3/2}\xi_{1i}\xi_{2a}\xi_{3b}
k_{3c}(k_{2d})\Tr(P_{-}\fsH_{(n)}M_p\Gamma^{bcaid}) (-t-s-u)L_1\nonumber\eeqa

which is exactly ${\cal A}_{1}$ of $<C^{-1}\phi^{-1}A^{0} A^{0}>$, where we have also used $p_d \epsilon^{a_{0}... a_{p-4}bcad}=0$, moreover without any further attempts, one reveals that  the first term  ${\cal A}_{4}$ of $<C^{-1}\phi^{0}A^{-1} A^{0}>$ is exactly 
the 1st term ${\cal A}_{4}$ of $<C^{-1}\phi^{-1}A^{0} A^{0}>$, however, in below we show that there is some other contact interaction that can just be found by 
$<C^{-1}\phi^{0}A^{-1} A^{0}>$ S-matrix.

\section {Other contact interaction to all orders in $\alpha'$}

From the direct computations we observed the fact that at the level of contact interactions  there is an extra contact term inside the S-matrix of $<C^{-1}\phi^{0}A^{-1} A^{0}>$ that has  been overlooked  from the direct calculations of $<C^{-1}\phi^{-1}A^{0} A^{0}>$ S-matrix. Indeed the  2nd contact interaction  ${\cal A}_{1}$ of   $<C^{-1}\phi^{0}A^{-1} A^{0}>$ is extra term  that is not only needed into the entire amplitude but also cannot be  derived from direct computations of $<C^{-1}\phi^{-1}A^{0} A^{0}>$ on upper half plane.

Hence the following coupling  is extra contact interaction to all orders in $\alpha'$ , which must have been appeared in S-matrix  because it stands correctly on the field theory side to all orders as well. Thus  let us first write it down

\beqa
4\pi^{1/2}\mu_p\xi_{1i}\xi_{2a}\xi_{3b} k_{3c}p^i\Tr(P_{-}\fsH_{(n)}M_p \Gamma^{bca})(-t-s-u)L_1\label{esiab}\eeqa
where we normalized the S-matrix by a coefficient of $(2\pi)^{1/2}\mu_p$. 



\vskip.1in

Having taken the expansion of $L_1$ inside \reef{esiab}, we first produce the leading term of string amplitude by the EFT coupling, in fact it can be produced by mixing Chern-Simons coupling and Taylor expansion as follows
 \beqa
S_3&=& \frac{i(2\pi\alpha')^3}{2}\mu_p\int d^{p+1}\sigma \quad \Tr(\prt_i C_{(p-3)}\wedge F \wedge F \Phi^i)
\label{s21}
\eeqa

Therefore one can explore the next order to the above coupling , which is $\alpha'^4$. Indeed all order $\alpha'$ corrections to the above coupling can be discovered by applying the proper higher derivative corrections on the above coupling  and the coefficients can just be fixed  by  taking the elements in the expansion of $L_1$ , so that all order corrections to above couplings are

\beqa
(s+t+u)^{n+1} H\phi A A&=&(\frac{\alpha'}{2})^{n+1}H (D_{a}D^{a})^{n+1}(\phi A A),\nonumber\\
(st)^{m}H\phi A A&=&(\alpha')^{2m}H D_{a_1}\cdots D_{a_{2m}}\Phi \partial^{a_{1}}\cdots \partial^{a_{m}}A
\partial^{a_{m+1}}\cdots \partial^{a_{2m}}A,\nonumber\\
(s+t)^{n}H\phi A A&=&(\alpha')^{n}H  D_{a_1}\cdots D_{a_{n}} \Phi \partial^{a_{1}}\cdots \partial^{a_{n}}(AA),\nonumber\\
(s)^{n}t^m H\phi A A&=&(\alpha')^{n+m}H D_{a_1}\cdots D_{a_{n}} D_{a_{1}}\cdots D_{a_{m}}\Phi \partial^{a_{1}}\cdots \partial^{a_{m}}A \partial^{a_{1}}\cdots \partial^{a_{n}}A,
\nonumber
\eeqa

   Note that in above couplings , inside the covariant derivative terms the  connections or commutator terms  do not appear and to get them, one needs to compute higher point amplitudes like $<C \phi AAA>$. 
   
   \vskip.1in

Therefore, we argue that for higher point function of string theory amplitudes, involving the mixed RR, a scalar and two gauge fields , there is a subtle issue as follows.

Indeed to be able to get to all the corrected and  all order contact interactions   as well as singularities  of the mixed  string theory amplitudes, one should  consider the scalar field  in zero picture as it has just been  clarified in detail by the comparisons of $<C^{-1}\phi^{0}A^{-1} A^{0}>$ with  $<C^{-1}\phi^{-1}A^{0} A^{0}>$  S-matrix.

 It would be nice to generalize this idea  to see  what happens for the mixed amplitudes of closed string RR,  two scalar fields  and one gauge field which we carry it out in the next section.
 
It would be even nicer if we could do it on asymmetric picture of RR ,that is,  find out $<C^{-2}\phi^{0}A^{0}A^{0}>$  to actually generalize the rules and symmetries of string theory, where we leave it for the future works, although partial results for simpler systems, like for brane anti brane have already been announced in \cite{Hatefi:2015gwa}.
Let us now  generate all the other singularity structures of $<C^{-1}\phi^{0}A^{-1}A^{0}>$ in the effective field theory.

\section { Other Singularities of $<C^{-1}\phi^{0}A^{-1} A^{0}>$}

Having produced some of the singularities and contact interactions,  we are now ready to derive some  other singularities of  $<C^{-1}\phi^{0}A^{-1} A^{0}>$ S-matrix. These singularities do exist for this five point world-sheet S-matrix which includes a symmetric RR, a scalar field in the zero picture and two gauge fields.
In fact by direct calculations we have shown that besides having some other contact interactions, even  at pole levels the whole ${\cal A}_{6}$ and ${\cal A}_{7}$ of   $<C^{-1}\phi^{0}A^{-1} A^{0}>$ are also extra singularities that cannot  be derived from the direct computations of $<C^{-1}\phi^{-1}A^{0} A^{0}>$ S-matrix. Let us write them down as follows

\beqa
{\cal A}_{7}&\sim&2^{-1/2} p^i \xi_{1i}\Tr(P_{-}\fsH_{(n)}M_p \gamma^{a})\xi_{2a} (2k_1.\xi_3) ut L_2\nonumber\\
{\cal A}_{6}&\sim&2^{-1/2}st L_2 p^i \xi_{1i}\Tr(P_{-}\fsH_{(n)}M_p \gamma^{b})
\bigg(2k_3.\xi_2 \xi_{3b}-2\xi_3.\xi_2 k_{3b}-2\xi_3.k_2\xi_{2b}\bigg)
\labell{48389}\eeqa

First we try to produce all these new s-channel poles of ${\cal A}_{7}$. By considering the desired expansion, one gets all  new s-channel poles (note that normalization constant is $(2\pi)^{1/2} m_p$) of $<C^{-1}\phi^{0}A^{-1} A^{0}>$ S-matrix as

\vskip.1in

\beqa
\frac{(2\pi\alpha')^2}{ p!}\mu_p p.\xi_1 \xi_{2a} 2k_1.\xi_3 \eps^{a_{0}\cdots a_{p-1}a}H_{a_{0}\cdots a_{p-1}}\sum_{n=-1}^{\infty}\frac{1}{s}{b_n(u+t)^{n+1}}
\Tr(\lam_1\lam_2\lam_3)\label{UI2}
\eeqa
In order to produce  all these new massless s-channel scalars , one has to apply the following field theory sub amplitude 
\beqa
{\cal A}&=&V^i_{\alpha}(C_{p-1},A_2,\phi)G^{ij}_{\alpha\beta}(\phi)V^j_{\beta}(\phi,A_3,\phi_1) \label{esi981}
\eeqa
where  to follow the related vertices, the kinetic term of scalar fields $ \frac{(2\pi\alpha')^2 }{2}D^a\phi^i D_a\phi_i$ has to be taken into account , so that we obtain
\beqa
V^j_{\beta}(\phi,A_3,\phi_1)&=&-2ik_1.\xi_3(2\pi\alpha')^2 T_p  \xi_1^j\Tr(\lam_3\lam_1\lam_\beta)
\labell{fvertex784}\\
({G}^{\phi})^{ij}_{\alpha\beta}&=&-\frac{i\delta^{ij}\delta^{\alpha\beta}}{(2\pi\alpha')^2T_p s}
\nonumber
\eeqa

Now one needs to re-consider the mixed Chern-Simons coupling and Taylor expended of scalar field, where the extremely important point has to be pointed out as follows. This turn, we take integration by parts and employ the momentum of external gauge field directly to RR (p-1) form potential to be able to produce the necessary field strength of RR whereas the total derivative terms are indeed zero at infinity, hence we find out the following effective action
 
\beqa
S_4&=&{i}(2\pi\alpha')^2\mu_p\int d^{p+1}\s {1\over (p-1)!}(\veps^v)^{a_0\cdots a_{p}}\prt_i H_{a_0\cdots a_{p-1}}
 \Tr\left(A_{a_{p}}\phi^i\right)\labell{interac38}
\eeqa
Having set the above action, we obtain the following vertex in the effective field theory 
\beqa
V^i_{\alpha}(C_{p-1},A_2,\phi)&=& p^i\frac{i (2\pi\alpha')^2\mu_p}{(p)!}(\veps^v)^{a_0\cdots a_p}H_{a_0\cdots a_{p-1}}\xi_{2a_{p}}\Tr(\lam_2\lambda_\alpha)
\eeqa

Note that $V^j_{\beta}(\phi,A_3,\phi_1)$ was  derived from the kinetic term of the scalar field and it has no correction , that is why to produce  all the singularities we need to propose all the  higher derivative corrections to the new action of \reef{interac38} as follows

\beqa
S_5&=&\sum_{n=-1}^{\infty}b_n (\alpha')^{(n+1)}\mu_p\int d^{p+1}\s {1\over (p-1)!}(\veps^v)^{a_0\cdots a_{p}}
\nonumber\\&&\times\partial_{i}H_{a_0\cdots a_{p-1}}D_{a_{1}}...D_{a_{n+1}} A_{a_{p}} D^{a_{1}}...D^{a_{n+1}}\phi^i
\label{vvcx89}
\eeqa

now we are allowed to actually reveal all order vertex operator of  $V^i_{\alpha}(C_{p-1},A_2,\phi)$ as

\beqa
V^i_{\alpha}(C_{p-1},A_2,\phi)&=& p^i \frac{i (2\pi\alpha')^2\mu_p}{(p)!}(\veps^v)^{a_0\cdots a_p}H_{a_0\cdots a_{p-1}}\xi_{2a_{p}}\Tr(\lam_2\lambda_\alpha)\sum_{n=-1}^{\infty}b_n(t+u)^{n+1}
\label{imb14}\eeqa

Replacing \reef{imb14} and \reef{fvertex784} to \reef{esi981}, we are then able to precisely regenerate  all order new s-channel singularities \reef{UI2} in the field theory side too. Finally let us reconstruct all new u-channel singularities.

\vskip.1in

Having replaced the desired expansion, we get all  new u-channel poles (normalisation constant is $(2\pi)^{1/2} m_p$) of string amplitude as follows

\beqa
\frac{(2\pi\alpha')^2}{ p!}\mu_p p.\xi_1 \eps^{a_{0}\cdots a_{p-1}b}H_{a_{0}\cdots a_{p-1}}\sum_{n=-1}^{\infty}\frac{1}{u}{b_n(s+t)^{n+1}}\bigg(2k_3.\xi_2 \xi_{3b}-2\xi_3.\xi_2 k_{3b}-2\xi_3.k_2\xi_{2b}\bigg)\label{UI26}
\eeqa

 All these u-channel gauge poles are also produced by considering the following sub amplitude in the field theory 

\beqa
{\cal A}&=&V^a_{\alpha}(C_{p-1},\phi_1,A)G^{ab}_{\alpha\beta}(A)V^b_{\beta}(A,A_2,A_3)\label{UI2678}
\eeqa

Here we consider the mixed Chern-Simons coupling and Taylor expended of scalar field, and not only this time  we take integration by parts but also we do apply the momentum of external gauge field directly to RR potential to be able to produce the necessary field strength of RR, 
keeping in mind the above remarks, we obtain the following vertex
\beqa
V^a_{\alpha}(C_{p-1},\phi_1,A)&=& p^i\frac{i (2\pi\alpha')^2\mu_p}{(p)!}(\veps^v)^{a_0\cdots a_{p-1}a}H_{a_0\cdots a_{p-1}}\xi_{1i}\Tr(\lam_1\lambda_\alpha)
\label{jip}\eeqa
where $V^b_{\beta}(A,A_2,A_3)$ has  no correction , so the only way of obtaining all the poles is to actually impose  all infinite higher derivative corrections to the mixed Chern-Simons Taylor expansion of scalar field,  so that now we can derive  the generalization of above vertex to all orders  as

\beqa
V^a_{\alpha}(C_{p-1},\phi_1,A)&=& p^i \frac{i (2\pi\alpha')^2\mu_p}{(p)!}(\veps^v)^{a_0\cdots a_{p-1}a}H_{a_0\cdots a_{p-1}}\xi_{1i}\Tr(\lam_1\lambda_\alpha)\sum_{n=-1}^{\infty}b_n(t+s)^{n+1}
\label{imb1o4}\eeqa

Now by taking into account  \reef{imb1o4}, the known   $V^b_{\beta}(A,A_2,A_3)$ and gauge field propagator 
$G_{\alpha\beta}^{ab}(A)=\frac{i\delta_{\alpha\beta}\delta^{ab}}{(2\pi\alpha')^2 T_p u}$ inside the sub amplitude \reef{UI2678}
 we are then able to precisely reconstruct  all order new u-channel singularities in the effective field theory side as well.


\vskip.1in


In the next section we further generalize our knowledge  by dealing with the mixed RR scalars/ gauge field S-matrices to see what happens to the S-matrix in the presence of two  scalar fields (in different pictures), a gauge field and a symmetric RR field strength.

\vskip.2in

\subsection{ All order S-matrix of   $<C ^{-1}A^{0}\phi ^{-1}\phi^{0}>$}

In this section we would like to see what is going on for the mixed higher point function of a  symmetric RR, two  transverse scalar fields (in two different pictures) and a gauge field. We do the whole details  to get to the entire  S-matrix to all orders in $\alpha'$  so the $<C ^{-1}A^{0}\phi ^{-1}\phi^{0}>$ S-matrix is shown by

\begin{eqnarray}
{\cal A}^{<C ^{-1}A^{0}\phi ^{-1}\phi^{0}>} & \sim & \int dx_{1}dx_{2}dx_{3}dzd\bar{z}\,
  \lan V_{A}^{(0)}{(x_{1})}
V_{\phi}^{(-1)}{(x_{2})}V_{\phi}^{(0)}{(x_{3})}
V_{RR}^{(-\frac{1}{2},-\frac{1}{2})}(z,\bar{z})\ran,\labell{sstring12678}\eeqa

Further simplification can be done to get to the closed form of S-matrix as follows

\beqa {\cal A}^{<C ^{-1}A^{0}\phi ^{-1}\phi^{0}>}&\sim& \int
 dx_{1}dx_{2}dx_{3}dx_{4} dx_{5}\,
(P_{-}\fsH_{(n)}M_p)^{\al\be}\xi_{1a}\xi_{2i}\xi_{3j}x_{45}^{-1/4}(x_{24}x_{25})^{-1/2}\nonumber\\&&
\times(I_1+I_2+I_3+I_4)\Tr(\lam_1\lam_2\lam_3),\labell{125678}\eeqa where
$x_{ij}=x_i-x_j$, $x_{4}=z$,$x_{5}=\bar z$, and also
\beqa
I_1&=&{<:\partial X^a(x_1)e^{\alpha' ik_1.X(x_1)}:e^{\alpha' ik_2.X(x_2)}
:\partial X^j(x_3)e^{\alpha' ik_3.X(x_3)}:e^{i\frac{\alpha'}{2}p.X(x_4)}:e^{i\frac{\alpha'}{2}p.D.X(x_5)}:>}
 \  \non \\&&\times{<:S_{\al}(x_4):S_{\be}(x_5):\psi^i(x_2):>},\nonumber\\
I_2&=&{<:\partial X^a(x_1)e^{\alpha' ik_1.X(x_1)}:e^{\alpha' ik_2.X(x_2)}
:e^{\alpha' ik_3.X(x_3)}:e^{i\frac{\alpha'}{2}p.X(x_4)}:e^{i\frac{\alpha'}{2}p.D.X(x_5)}:>}
 \  \non \\&&\times{<:S_{\al}(x_4):S_{\be}(x_5)::\psi^i(x_2):\alpha' ik_{3c}\psi^{c}\psi^{j}(x_3)>},\nonumber\\
 I_3&=&{<: e^{\alpha' ik_1.X(x_1)}:e^{\alpha' ik_2.X(x_2)}
:\partial X^j(x_3)e^{\alpha' ik_3.X(x_3)}:e^{i\frac{\alpha'}{2}p.X(x_4)}:e^{i\frac{\alpha'}{2}p.D.X(x_5)}:>}
 \  \non \\&&\times{<:S_{\al}(x_4):S_{\be}(x_5):\alpha' ik_{1b}\psi^{b}\psi^{a}(x_1):\psi^i(x_2):>},\nonumber\\
 I_4&=&{<: e^{\alpha' ik_1.X(x_1)}:e^{\alpha' ik_2.X(x_2)}
:e^{\alpha' ik_3.X(x_3)}:e^{i\frac{\alpha'}{2}p.X(x_4)}:e^{i\frac{\alpha'}{2}p.D.X(x_5)}:>}
 \  \non \\&&\times{<:S_{\al}(x_4):S_{\be}(x_5):\alpha' ik_{1b}\psi^{b}\psi^a(x_1):\psi^i(x_2)
:\alpha' ik_{3c}\psi^{c}\psi^j(x_3):>}.
\label{i1234675}
\eeqa

\vskip 0.1in

If we work with  all possible contractions, then one finds out the compact form of the following  fermionic correlation function as follows

\beqa
I_6^{jciab}&=&<:S_{\al}(x_4):S_{\be}(x_5):\psi^b\psi^a(x_1)::\psi^i(x_2):\psi^c\psi^j(x_3)>\nonumber\\
&=&\bigg\{(\Gamma^{jciab}C^{-1})_{{\alpha\beta}}+\alpha' r_1\frac{Re[x_{14}x_{35}]}{x_{13}x_{45}}+\alpha' r_2\frac{Re[x_{24}x_{35}]}{x_{23}x_{45}}\nonumber\\&&+\alpha'^2 r_3(\frac{Re[x_{14}x_{35}]}{x_{13}x_{45}})(\frac{Re[x_{24}x_{35}]}{x_{23}x_{45}})
\bigg\}2^{-5/2}x_{45}^{5/4}(x_{14}x_{15}x_{34}x_{35})^{-1}(x_{24}x_{25})^{-1/2},
\label{hh11657}
\eeqa
where
\beqa
r_1&=&\bigg(-\eta^{bc}(\Gamma^{jia}C^{-1})_{\alpha\beta}+\eta^{ac}(\Gamma^{jib}C^{-1})_{\alpha\beta}\bigg),\nonumber\\
r_2&=&\bigg(\eta^{ij}(\Gamma^{cab}C^{-1})_{\alpha\beta}\bigg),\nonumber\\
r_3&=&\bigg(\eta^{bc}\eta^{ij}(\gamma^{a}C^{-1})_{\alpha\beta}-\eta^{ac}\eta^{ij}(\gamma^{b}C^{-1})_{\alpha\beta}\bigg)
\eeqa
Substituting  the closed form of the correlators into the amplitude we now claim the final answer for the S-matrix can be written down by
\beqa
{\cal A}^{<C ^{-1}A^{0}\phi ^{-1}\phi^{0}>}&\!\!\!\!\sim\!\!\!\!\!&\int dx_{1}dx_{2} dx_{3}dx_{4}dx_{5}(P_{-}\fsH_{(n)}M_p)^{\al\be}I\xi_{1a}\xi_{2i}\xi_{3j}x_{45}^{-1/4}(x_{24}x_{25})^{-1/2}\nonumber\\&&\times
\bigg(I_7^i( a^a_1a^j_3)+a^a_1a^{ji}_2+a^j_3a^{ia}_4-\alpha'^2 k_{1b}k_{3c}I_6^{jciab}\bigg)\Tr(\lam_1\lam_2\lam_3)
\labell{amp31q657},\eeqa
where
\beqa
I&=&|x_{12}|^{\alpha'^2 k_1.k_2}|x_{13}|^{\alpha'^2 k_1.k_3}|x_{14}x_{15}|^{\frac{\alpha'^2}{2} k_1.p}|x_{23}|^{\alpha'^2 k_2.k_3}|
x_{24}x_{25}|^{\frac{\alpha'^2}{2} k_2.p}
|x_{34}x_{35}|^{\frac{\alpha'^2}{2} k_3.p}|x_{45}|^{\frac{\alpha'^2}{4}p.D.p},\nonumber\\
a^a_1&=&ik_2^{a}\bigg(\frac{x_{42}}{x_{14}x_{12}}+\frac{x_{52}}{x_{15}x_{12}}\bigg)+ik_3^{a}\bigg(\frac{x_{43}}{x_{14}x_{13}}+\frac{x_{53}}{x_{15}x_{13}}\bigg)\nonumber\\
a^j_3&=&ip^{j}\bigg(\frac{x_{54}}{x_{34}x_{35}}\bigg)\nonumber\\
a^{ji}_2&=& \bigg\{(\Gamma^{jci}C^{-1})_{\alpha\beta}+(\alpha'\eta^{ij}(\gamma^{c}C^{-1})_{\alpha\beta})\frac{Re[x_{24}x_{35}]}{x_{23}x_{45}}\bigg\}\nonumber\\&&\times
\alpha' ik_{3c}2^{-3/2}x_{45}^{1/4}(x_{34}x_{35})^{-1}(x_{24}x_{25})^{-1/2}
\nonumber\\
a^{ia}_4&=&\alpha' ik_{1b}2^{-3/2}x_{45}^{1/4}(x_{24}x_{25})^{-1/2}(x_{14}x_{15})^{-1} \bigg\{(\Gamma^{iab}C^{-1})_
{\alpha\beta}\bigg\}
,\nonumber\\
I_7^i&=&<:S_{\al}(x_4):S_{\be}(x_5):\psi^i(x_2):>=2^{-1/2}x_{45}^{-3/4}(x_{24}x_{25})^{-1/2}
(\gamma^{i}C^{-1})_{\alpha\beta}.\nonumber
\eeqa

 \vskip 0.2in

It now becomes  clear that the S-matrix of \reef{amp31q657} is SL(2,R) invariant and after gauge fixing over the position of open strings one needs to come over the integrals on upper half complex plane on the location of RR. By  evaluating those  integrals one eventually writes down the complete form of the S-matrix to all orders as follows
\beqa {\cal A}^{<C ^{-1}A^{0}\phi ^{-1}\phi^{0}>}&=&{\cal A}_{1}+{\cal A}_{2}+{\cal A}_{3}+{\cal A}_{4}+{\cal A}_{5}
+{\cal A}_{6}
\labell{71112u}\eeqa
where
\beqa
{\cal A}_{1}&\!\!\!\sim\!\!\!&2^{-1/2}\xi_{1a}\xi_{2i}\xi_{3j} p^j\Tr(P_{-}\fsH_{(n)}M_p\gamma^{i})
\bigg[-2k^a_3(ut)+2k^a_2(us)\bigg]L_2
\nonumber\\
{\cal A}_{2}&\sim&2^{-1/2}k_{3c}\bigg\{-2k_2.\xi_1 \xi_{2i}\xi_{3j}(us)L_2\Tr(P_{-}\fsH_{(n)}M_p \Gamma^{jci})
+2k_3.\xi_1 \xi_{2i}\xi_{3j}(ut)L_2\Tr(P_{-}\fsH_{(n)}M_p \Gamma^{jci})
\nonumber\\&&+4t\xi_2.\xi_3 k_3.\xi_1L_1\Tr(P_{-}\fsH_{(n)}M_p \gamma^{c})
-4s\xi_2.\xi_3 k_2.\xi_1L_1\Tr(P_{-}\fsH_{(n)}M_p \gamma^{c})\bigg\}\nonumber\\
{\cal A}_{3}&\sim&2^{-1/2} k_{1b}\xi_{1a}\xi_{2i}\xi_{3j}4(-u-s-t)L_1\bigg(\Tr(P_{-}\fsH_{(n)}M_p\Gamma^{iab}) p^j-k_{3c}\Tr(P_{-}\fsH_{(n)}M_p\Gamma^{jciab})\bigg)\nonumber\\
{\cal A}_{4}&\sim&2^{-1/2}(ut)L_2
\bigg\{ -s\xi_{1a}\xi_{2i}\xi_{3j}\Tr(P_{-}\fsH_{(n)}M_p \Gamma^{jia})-2k_3.\xi_1 k_{1b}\xi_{2i}\xi_{3j}\Tr(P_{-}\fsH_{(n)}M_p \Gamma^{jib})\bigg\}\nonumber\\
{\cal A}_{5}&\sim&2^{1/2}(st)L_2\xi_2.\xi_3\xi_{1a}k_{1b}k_{3c}\Tr(P_{-}\fsH_{(n)}M_p \Gamma^{cab})\nonumber\\
{\cal A}_{6}&\sim&2^{1/2}\xi_{3}.\xi_{2}\bigg(ts\Tr(P_{-}\fsH_{(n)}M_p \gamma^{a})\xi_{1a}+2tk_3.\xi_1
\Tr(P_{-}\fsH_{(n)}M_p \gamma^{b})k_{1b}\bigg)L_1
\labell{483765}\eeqa
where the functions
 $L_1,L_2$ are given in \reef{Ls2345}.

\vskip.3in

On the other hand if we actually consider both scalar fields in zero picture in the presence of a symmetric RR, then  we get the whole S-matrix as

 \begin{eqnarray}
{\cal A}^{<C^{-1}A^{-1}\phi^{0}\phi^{0}>} & \sim & \int dx_{1}dx_{2}dx_{3}dzd\bar{z}\,
  \lan V_{A}^{(-1)}{(x_{1})}
V_{\phi}^{(0)}{(x_{2})}V_\phi^{(0)}{(x_{3})}
V_{RR}^{(-\frac{1}{2},-\frac{1}{2})}(z,\bar{z})\ran,\nonumber\eeqa

Having done all integrals, one could find the final answer ( for further details , look at \cite{Hatefi:2012zh})
 for the entire S-matrix of a  symmetric RR with both transverse scalars in zero picture and a gauge field as follows
\beqa {\cal A}^{<C^{-1}A^{-1}\phi^{0}\phi^{0}>}&=&{\cal A}_{1}+{\cal A}_{2}+{\cal A}_{3}+{\cal A}_{4}+{\cal A}_{5}+{\cal A}_{6}
+{\cal A}_{7}+{\cal A}_{8}+{\cal A}_{9}+{\cal A}_{10}\labell{11u}\eeqa
where
\beqa
{\cal A}_{1}&\!\!\!\sim\!\!\!&-2^{-1/2}\xi_{1a}\xi_{2i}\xi_{3j}
\bigg[k_{3c}k_{2b}\Tr(P_{-}\fsH_{(n)}M_p\Gamma^{jciba})-k_{2b}p^j\Tr(P_{-}\fsH_{(n)}M_p\Gamma^{iba})\nonumber\\&&-k_{3c}p^i\Tr(P_{-}\fsH_{(n)}M_p\Gamma^{jca})+p^ip^j\Tr(P_{-}\fsH_{(n)}M_p\gamma^{a})\bigg]
4(-s-t-u)L_1,
\nonumber\\
{\cal A}_{2}&\sim&2^{-1/2}
\bigg\{-
2\xi_{1}.k_{2}k_{3c}\xi_{3j}\xi_{2i}\Tr(P_{-}\fsH_{(n)}M_p \Gamma^{jci})\bigg\}(us)L_2\nonumber\\
{\cal A}_{3}&\sim&2^{-1/2}
\bigg\{\xi_{1a}\xi_{2i}\xi_{3j}\Tr(P_{-}\fsH_{(n)}M_p \Gamma^{jia})\bigg\}(-ust)L_2\nonumber\\
{\cal A}_{4}&\sim&2^{-1/2}
\bigg\{
2k_{3}.\xi_{1}k_{2b}\xi_{3j}\xi_{2i}\Tr(P_{-}\fsH_{(n)}M_p \Gamma^{jib})\bigg\}(ut)L_2\nonumber\\
\nonumber\\
{\cal A}_{5}&\sim&2^{-1/2}
\bigg\{2\xi_{3}.\xi_{2}k_{2b}k_{3c}\xi_{1a}\Tr(P_{-}\fsH_{(n)}M_p \Gamma^{cba})\bigg\}(st) L_2\nonumber\\
{\cal A}_{6}&\sim& 2^{1/2}(us) L_{2}\bigg\{p^j\xi_1.k_2\xi_{2i}\xi_{3j}\Tr(P_{-}\fsH_{(n)}M_p\gamma^i)
\bigg\}
\nonumber\\
{\cal A}_{7}&\sim&-2^{-1/2} (ut) L_2\bigg\{
2k_{3}.\xi_1p^i\xi_{3j}\xi_{2i}\Tr(P_{-}\fsH_{(n)}M_p\gamma^j)\bigg\}
\nonumber\\
{\cal A}_{8}&\sim&2^{1/2}L_1\bigg\{2k_2.\xi_1 k_{3c}\Tr(P_{-}\fsH_{(n)}M_p\gamma^c)
(-s\xi_2.\xi_3)\bigg\}.
\nonumber\\
{\cal A}_{9}&\sim&2^{1/2}L_1\bigg\{2k_3.\xi_{1}k_{2b}\Tr(P_{-}\fsH_{(n)}M_p\gamma^b)
(-t\xi_2.\xi_3)\bigg\}
\nonumber\\
{\cal A}_{10}&\sim&2^{1/2}L_1\bigg\{\xi_{1a}\Tr(P_{-}\fsH_{(n)}M_p\gamma^a)
(ts\xi_3.\xi_2)\bigg\}
\labell{480}\eeqa
where the functions
 $L_1,L_2$ are already appeared in \reef{Ls2345}.

\vskip 0.2in

It is worth highlighting the point that, this S-matrix also satisfies Ward identity, that is, by substituting $\xi_{1a}\rightarrow k_{1a}$, the entire amplitude vanishes and the amplitude holds for various $p,n$ cases. Let us  do the comparisons $<C^{-1}A^{0}\phi^{-1}\phi^{0}>$ with $<C^{-1}A^{-1}\phi^{0}\phi^{0}>$ S-matrix at both level of singularity structures and contact interactions, find out various new couplings and in particular find out their corrections and eventually  get to the conclusion.

\section{Comparison on Singularity Structure of $<C^{-1}A^{0}\phi^{-1}\phi^{0}>$ with $<C^{-1}A^{-1}\phi^{0}\phi^{0}>$ }

In this section we are going to compare all the  singularities of $<C^{-1}A^{0}\phi^{-1}\phi^{0}>$ with $<C^{-1}A^{-1}\phi^{0}\phi^{0}>$ S-matrix. The first term  ${\cal A}_{6}$ of $<C^{-1}A^{0}\phi^{-1}\phi^{0}>$ is exactly equivalent to  ${\cal A}_{10}$ of $<C^{-1}A^{-1}\phi^{0}\phi^{0}>$, likewise the last term  ${\cal A}_{2}$ of $<C^{-1}A^{0}\phi^{-1}\phi^{0}>$ is the same as
${\cal A}_{8}$ of $<C^{-1}A^{-1}\phi^{0}\phi^{0}>$ S-matrix.

\vskip 0.2in

Now if we add the second term  ${\cal A}_{6}$ of $<C^{-1}A^{0}\phi^{-1}\phi^{0}>$
with the third term  ${\cal A}_{2}$ of $<C^{-1}A^{0}\phi^{-1}\phi^{0}>$ and make use of momentum conservation along the world volume of branes, we obtain

\beqa
2^{1/2} L_1 (2tk_3.\xi_1)\xi_{2}.\xi_{3}\Tr(P_{-}\fsH_{(n)}M_p\gamma^{b}) (-k_{2b}-p_{b})\nonumber\eeqa

Now by applying the following equation    $p_b \epsilon^{a_{0}... a_{p-1}b}=0$, we then realize the fact that  the first term in above equation precisely produces the ${\cal A}_{9}$ term of
$<C^{-1}A^{-1}\phi^{0}\phi^{0}>$.

Meanwhile ${\cal A}_{5}$ of $<C^{-1}A^{0}\phi^{-1}\phi^{0}>$ can be written down as
\beqa
2^{1/2} (st)L_2  \xi_{1a}\xi_{2}.\xi_{3}k_{3c}\Tr(P_{-}\fsH_{(n)}M_p\Gamma^{cab}) (-k_{3b}-k_{2b}-p_{b})\nonumber\eeqa

where  the first term has no contribution to S-matrix. Because of the antisymmetric property of $\epsilon$  and the fact that it is symmetric with respect to $k_3$ so the result for the  first term is zero. More evidently the third term in above equation has no contribution because  $p_b \epsilon^{a_{0}... a_{p-3}cab}=0$ and the second term precisely produces ${\cal A}_{5}$ of $<C^{-1}A^{-1}\phi^{0}\phi^{0}>$.

 \vskip.2in

 The same so happens to the other terms, namely if we add the 2nd terms of ${\cal A}_{2}$ and
${\cal A}_{4}$ of $<C^{-1}A^{0}\phi^{-1}\phi^{0}>$ and apply the momentum conservation, then we are able to precisely produce ${\cal A}_{4}$ of $<C^{-1}A^{-1}\phi^{0}\phi^{0}>$ which is related to all s-channel poles.

 \vskip.1in

Indeed without any further details the first term  ${\cal A}_{2}$ of $<C^{-1}A^{0}\phi^{-1}\phi^{0}>$ is exactly ${\cal A}_{2}$
term of $<C^{-1}A^{-1}\phi^{0}\phi^{0}>$ so that all  t-channel poles are then reproduced in both pictures. By considering the 2nd term ${\cal A}_{1}$ of $<C^{-1}A^{0}\phi^{-1}\phi^{0}>$ we are then able to generate ${\cal A}_{6}$
term of $<C^{-1}A^{-1}\phi^{0}\phi^{0}>$.

\vskip.2in

Finally  to be able to produce all  the second kind of s-channel poles one has to  subtract the first term of ${\cal A}_{1}$ of $<C^{-1}A^{0}\phi^{-1}\phi^{0}>$ from ${\cal A}_{7}$
term of $<C^{-1}A^{-1}\phi^{0}\phi^{0}>$ such that upon considering the following identity

\beqa
\xi_{2}\xi_{3j} \epsilon^{a_{0}...a_{p}}(-p^j H^{i}_{a_{0}...a_{p}}+p^i H^{j}_{a_{0}...a_{p}})&=&0
\nonumber\eeqa

we believe that the first term of ${\cal A}_{1}$ of $<C^{-1}A^{0}\phi^{-1}\phi^{0}>$ is exactly the same  ${\cal A}_{7}$
term of $<C^{-1}A^{-1}\phi^{0}\phi^{0}>$.

\vskip.2in
Henceforth,  we could precisely produce all the singularities of this five point function in two different pictures. However, note that  we have some  extra  contact interactions in  $<C^{-1}A^{-1}\phi^{0}\phi^{0}>$  amplitude while they are absent in $<C^{-1}A^{0}\phi^{-1}\phi^{0}>$ S-matrix.  These extra contact interactions are needed by symmetries of string theory amplitudes as we point out/hint them in a moment.

 \vskip.2in

For the completeness we first would like to produce all the singularities. This amplitude has  u-channel  gauge poles that can be read off  from the string amplitude as follows 
\beqa
\mu_p(2\pi\alpha')^{2} 2k_{2a} k_{3a_{p-1}}\xi_2.\xi_3\frac{1}{(p-2)!u}\eps^{a_{0}\cdots a_{p-1}a}H_{a_{0}\cdots a_{p-3}}\xi_{1a_{p-2}}\sum_{n=-1}^{\infty}b_n\bigg(\frac{\alpha'}{2}\bigg)^{n+1}(s+t)^{n+1}
\label{ope23}\eeqa

where these u-channel poles should be produced by the following sub amplitude in the effective field theory
\beqa
{\cal A}&=&V^a_{\alpha}(C_{p-3},A_1,A)G^{ab}_{\alpha\beta}(A)V^b_{\beta}(A,\phi_2,\phi_3),\label{amp644}
\eeqa







Considering the kinetic terms of scalars $iT_p \frac{(2\pi\alpha')^2}{2}\Tr(D^a\phi^iD_a\phi_i)$ and gauge fields we  obtain  the following vertices
\beqa
V_{\beta}^{b}(A,\phi_2,\phi_3)&=&i\lambda^2T_p
  \xi_2.\xi_3 (k_2-k_3)^b \Tr(\lambda_2\lambda_3\lambda_\beta)
\nonumber\\
G_{\alpha\beta}^{ab}(A)&=&\frac{-i}{\lambda^2T_p}\frac{\delta^{ab}
\delta_{\alpha\beta}}{k^2}\,\,\, ,
\label{ver137}
\eeqa



The kinetic terms have no corrections so we need to apply all higher derivative  corrections to Chern-Simons couplings as follows
 \beqa
 S_6&=& i(2\pi\alpha')^2\mu_p\int d^{p+1}\sigma   \quad \sum_{n=-1}^{\infty}b_n (\alpha')^{n+1}\quad C_{(p-3)}\wedge D_{a_0\cdots a_n} F \wedge D^{a_0\cdots a_n} F
\eeqa
Now if one considers $S_6$, then one is able to obtain the following vertex operator to all orders in $\alpha'$ as follows
 \beqa
V^a_{\alpha}(C_{p-3},A_1,A)&=&\frac{\lambda^2\mu_p}{(p-2)!}(\eps)^{a_0\cdots a_{p-1}a}(H^{(p-2)})_{a_0\cdots a_{p-3}}\xi_{1a_{p-2}}k_{a_{p-1}}\nonumber\\&&\times
\Tr(\lam_1\lambda_\alpha)\sum_{n=-1}^{\infty}b_n(\alpha'k_1.k)^{n+1}
\label{990}\eeqa

Replacing above vertices \reef{990} and \reef{ver137} into \reef{amp644}, we are then able to exactly produce all  u-channel gauge poles in the field theory side.


\vskip.1in

On the other hand, if we employ  all order $\alpha'$ SYM couplings as appeared in \reef{highder}, 

and also  apply a following sub amplitude of field theory 

\beqa
{\cal A}&=&V_{\alpha}^{a}(C_{p-1},A)G_{\alpha\beta}^{ab}(A)V_{\beta}^{b}(A,A_1,\phi_2,\phi_3)\nonumber\eeqa
 then we will be able to produce all   $(t+s+u)$ gauge field poles. Note that  this task has been completely done in section four  of \cite{Hatefi:2012zh} and in order to avoid rewriting the old contents of the paper, we refer the interested reader to  that section four of \cite{Hatefi:2012zh}.

\vskip.1in

Let us reconstruct all  t-channel poles and finally by interchanging $1\leftrightarrow 2$ for all the momenta, the polarisations and t to s,  we are able to produce all s-channel poles as well. All the t-channel poles of the string amplitude are given by

 \beqa
 \frac{16\xi_{2i}\xi_{3j}k_2.\xi_1\pi^2 \mu_p}{t(p+1)!}
\bigg\{2 p^j\eps^{a_{0}\cdots a_{p}}H^{i}_{a_{0}\cdots a_{p}}
-2(p+1)k_{3a}\eps^{a_{0}\cdots a_{p-1}a}H^{ij}_{a_{0}\cdots a_{p-1}}
\bigg\}\sum_{n=-1}^{\infty} b_n (\alpha'k_3.k)^n
\label{connor}\eeqa
These t-channel poles can be regenerated in the field theory side, and to do so one needs to take into account the following sub amplitude and vertices in the field theory as 

\beqa
{\cal A}&=&V^i_{\alpha}(C_{p+1},\phi_3,\phi)G^{ij}_{\alpha\beta}(\phi)V^j_{\beta}(\phi,A_1,\phi_2)\nonumber\\
V^j_{\beta}(\phi,A_1,\phi_2)&=&-2i(2\pi\alpha')^2T_p k_2.\xi_1
\xi^j_2 \Tr(\lambda_1\lambda_2\lambda_\beta)
\nonumber\\
G_{\alpha\beta}^{ij}(\phi)&=&\frac{-i}{(2\pi\alpha')^2T_p}\frac{\delta^{ij}
\delta_{\alpha\beta}}{t}
\label{ver13}
\eeqa

 Consider the Taylor expansion of the two scalar fields as 
\beqa
S_{7}&=&\frac{(2\pi\alpha')^2\mu_p}{2}\int d^{p+1}\sigma \frac{1}{(p+1)!}\eps^{a_0\cdots a_{p}}
  \Tr\left(\Phi^j \Phi^i\right)
\prt_j\prt_iC^{(p+1)}_{a_0\cdots a_{p}}\nonumber\eeqa

and then work out with pull-back and both mixing term involving  Taylor and pull-back as follows
\beqa
S_{8}
&=&\frac{(2\pi\alpha')^2\mu_p}{2}\int d^{p+1}\sigma \frac{1}{(p+1)!}
\eps^{a_0\cdots a_{p}}\bigg[p(p+1)\,
\Tr\left(D_{a_0}\Phi^i\,D_{a_1}\Phi^j\right)
C^{(p+1)}_{ija_2\cdots a_{p}}\nonumber\\&&+2(p+1)
\Tr\left(\Phi^j D_{a_0}\Phi^i\right)
\prt_jC^{(p+1)}_{ia_1\cdots a_{p}}
\bigg]
\label{221}
\eeqa

where one needs to also add the following Myers terms 

\beqa
S_{9}&=&{i\over4}(2\pi\alpha')^2\mu_p\int d^{p+1}\sigma {1\over(p-1)!} \eps^{a_0\cdots a_{p}}
\,\Tr\left(F_{a_0a_1}[\Phi^j,\Phi^i]\right)
C^{(p+1)}_{ija_2\cdots a_{p}} .
\label{6733}
\eeqa

 with $S_{8}$ and take all the integrations by parts to actually get to the following action

\beqa
S_{10}&=&{(2\pi\alpha')^2\over2}\mu_p\int d^{p+1}\sigma {1\over (p+1)!}
\eps^{a_0\cdots a_{p}}\left[(p+1)\Tr\left(D_{a_0}\Phi^j\Phi^i\right)
H^{(p+2)}_{ija_1\cdots a_{p}}\right]
\nonumber\eeqa

Eventually  in order to produce the  first  t-channel pole, one must consider the summation of the Taylor expansion and $S_{10}$ as follows

\beqa
\frac{\mu_p(2\pi\alpha')^2}{2(p+1)!}\int d^{p+1}\sigma
\eps^{a_0\cdots a_{p}}\left[\Tr\bigg(\Phi^j\Phi^i\bigg)
\prt_j H^{(p+2)}_{ia_0\cdots a_{p}}
+(p+1)\Tr\left(D_{a_0}\Phi^j\Phi^i\right)
H^{(p+2)}_{ija_1\cdots a_{p}}\right]
\label{5yghtt}
\eeqa

From \reef{5yghtt} we now look for the vertex of $V^i_{\alpha}(C_{p+1},\phi_3,\phi)$ as follows
\beqa
V^i_{\alpha}(C_{p+1},\phi_3,\phi)&=&\frac{\mu_p(2\pi\alpha')^2}{(p+1)!}\Tr(\lambda_3\lambda_{\alpha})
\eps^{a_0\cdots a_{p}}\bigg[p^j \xi_{3j}H^{i}_{a_0\cdots a_{p}}
+(p+1)H^{ij}_{a_1\cdots a_{p}}k_{3a_{0}} \xi_{3j} \bigg]
\label{ppo}\eeqa

However, to produce all the other t-channel poles, one needs to apply all order higher derivative corrections to \reef{5yghtt} as below  

\beqa
&&\frac{\mu_p(2\pi\alpha')^2}{2(p+1)!}\int d^{p+1}\sigma
\eps^{a_0\cdots a_{p}} \sum_{n=-1}^{\infty} b_n (\alpha')^n \bigg[\Tr\bigg(D_{a_{1}...a_{n}}\Phi^jD^{a_{1}...a_{n}}\Phi^i\bigg)
\prt_j H^{(p+2)}_{ia_0\cdots a_{p}}\nonumber\\&&
+(p+1)\Tr\left(D_{a_0}D_{a_{1}...a_{n}}\Phi^jD^{a_{1}...a_{n}}\Phi^i\right)
H^{(p+2)}_{ija_1\cdots a_{p}}\bigg]
\label{5gh}
\eeqa

to indeed obtain the following vertex to all orders in $\alpha'$ as follows

  \beqa
V^i_{\alpha}(C_{p+1},\phi_3,\phi)&=&\frac{\mu_p(2\pi\alpha')^2}{(p+1)!}\Tr(\lambda_3\lambda_{\alpha})
\eps^{a_0\cdots a_{p}}\sum_{n=-1}^{\infty} b_n (\alpha'k_3.k)^n\nonumber\\&&\times\bigg[p^j \xi_{3j}H^{i}_{a_0\cdots a_{p}}
+(p+1)H^{ij}_{a_1\cdots a_{p}}k_{3a_{0}} \xi_{3j} \bigg]
\label{ppo12}\eeqa
Now if we replace \reef{ppo12} inside \reef{ver13} then we are exactly able to regenerate all order t-channel singularities in the field theory side as well.

Note that all of  the new couplings that we have discovered, can just be derived with scattering computations not by any duality transformation. Because the  coefficients of these couplings can just be fixed without any ambiguity by S-matrix analysis. We now turn to contact interaction terms.

\section{Comparison on  Contact interactions }

 If we look at the precise computations of the S-matrices in two different pictures,  we then  realize the fact that the first term  
 ${\cal A}_{4}$ of $<C^{-1}A^{0}\phi^{-1}\phi^{0}>$ is exactly the term that has been shown up in ${\cal A}_{3}$ of $<C^{-1}A^{-1}\phi^{0}\phi^{0}>$.

 As we can readily observe, we have just left with two contact terms in ${\cal A}_{3}$
 of $<C^{-1}A^{0}\phi^{-1}\phi^{0}>$ while in  ${\cal A}_{1}$ of $<C^{-1}A^{-1}\phi^{0}\phi^{0}>$ we do have four different terms, so let us keep comparing.

 Now if we apply the momentum conservation to the 2nd term  ${\cal A}_{3}$
 of $<C^{-1}A^{0}\phi^{-1}\phi^{0}>$  and apply the Bianchi equation that we have already got , that is,  $p_b \epsilon^{a_{0}..a_{p-3}cba}=0$ then we are able to precisely produce the first term ${\cal A}_{1}$ of $<C^{-1}A^{-1}\phi^{0}\phi^{0}>$.

 \vskip.2in

Eventually  we apply momentum conservation to  the only remaining term of $<C^{-1}A^{0}\phi^{-1}\phi^{0}>$ which is its first ${\cal A}_{3}$ term and do  subtract it from the second and third terms  ${\cal A}_{1}$
of  $<C^{-1}A^{-1}\phi^{0}\phi^{0}>$ such that upon holding the  following equation, we are able to generate the second and third term 
 ${\cal A}_{1}$ of  $<C^{-1}A^{-1}\phi^{0}\phi^{0}>$. 

\beqa
\xi_{2i}\xi_{3j}\xi_{1a}k_{3b}
\epsilon^{a_{0}..a_{p-2}ab}(p^j H^{i}_{a_{0}..a_{p-2}}-p^i H^{j}_{a_{0}..a_{p-2}})&=&0
\nonumber\eeqa

Once more $p_b \epsilon^{a_{0}..a_{p-2}ab}=0$, whereas up to a sign the third term   ${\cal A}_{1}$
of  $<C^{-1}A^{-1}\phi^{0}\phi^{0}>$ is also produced.

However, note to the important point that there is no chance to actually produce even the leading order $\alpha'$ of the fourth contact interaction  ${\cal A}_{1}$
of  $<C^{-1}A^{-1}\phi^{0}\phi^{0}>$. The reason is that, there is no left over term inside $<C^{-1}A^{0}\phi^{-1}\phi^{0}>$ S-matrix to be compared  with  that fourth term ${\cal A}_{1}$
of  $<C^{-1}A^{-1}\phi^{0}\phi^{0}>$ S-matrix. Therefore, let us further elaborate on the needed contact interactions of this string amplitude.
 

\section {The needed contact interaction for $<C^{-1}\phi^{-1}\phi^{0} A^{0}>$}

As we have seen above, we were able to produce all the first three contact terms ${\cal A}_{1}$
of  $<C^{-1}A^{-1}\phi^{0}\phi^{0}>$ to all orders, however, we have evidently observed that indeed there is no chance to produce the fourth term  contact interaction  ${\cal A}_{1}$ of  $<C^{-1}A^{-1}\phi^{0}\phi^{0}>$ by direct computations of $<C^{-1}A^{0}\phi^{-1}\phi^{0}>$. 

In fact we claim that this extra contact interaction must be appeared in the entire S-matrix as it plays the crucial role in all order $\alpha'$ contact interaction terms in both type IIA and IIB  super string theory. Let us first write it down and then we try to construct its all order $\alpha'$ higher derivative couplings.

\vskip.1in

Hence we figure out the  following term inside $<C^{-1}A^{-1}\phi^{0}\phi^{0}>$ S-matrix

 \beqa
-4\pi^{1/2}\mu_p\xi_{1a}\xi_{2i}\xi_{3j} p^i p^j\Tr(P_{-}\fsH_{(n)}M_p \gamma^{a}) (-t-s-u) L_1\label{esi12}\eeqa

 is indeed needed. We normalized the S-matrix by a coefficient of $(2\pi)^{1/2}\mu_p$ and considered the expansion of $L_1$  
(with the aforementioned coefficients) given in \reef{highcaap}.  Thus  the first leading term of $L_1$ can be produced by  Chern-Simons coupling and Taylor expanded of both scalar fields through closed string RR as follows
 \beqa
S_{11}&=& \frac{i(2\pi\alpha')^3}{2}\mu_p\int d^{p+1}\sigma \quad \Tr(\prt_j\prt_i C_{(p-1)}\wedge F  \Phi^i\Phi^j)
\label{s24r1}
\eeqa

Therefore one  explores the next order term which is $\alpha'^4$ and indeed all order $\alpha'$ corrections to the above coupling with exact coefficients can be discovered by  applying the proper higher derivative corrections. For example the $(st)^{m}H A\phi \phi$ and $(s+t)^{n}H A\phi \phi$ contact terms of the S-matrix (inside the expansion of $L_1$)
 can be shown to be matched to all orders by the following couplings
\beqa
(s+t)^{n}H A\Phi \Phi&=&(\alpha')^{n}H  \partial_{a_1}\cdots \partial_{a_{n}} A D^{a_{1}}\cdots D^{a_{n}}(\Phi\Phi),\nonumber\\
(st)^{m}H A\Phi \Phi&=&(\alpha')^{2m}H \partial_{a_1}\cdots \partial_{a_{2m}}A D^{a_{1}}\cdots D^{a_{m}}\Phi
D^{a_{m+1}}\cdots D^{a_{2m}}\Phi\nonumber
\eeqa

    Note that the first correction to the above coupling \reef{s24r1} and the other new coupling in \reef{s21} is  of $\alpha'^4$ order.

\vskip.2in

It is also worth keeping in mind  the fact that by  expanding the string amplitude of $<C^{-1}A^{-1}\phi^{0}\phi^{0}>$,  we could also explore new couplings at leading order as follows.

\vskip.1in


Let us write down the   explicit form of the string amplitude, indeed if we extract the related trace, consider the expansion of $st L_2$ inside ${\cal A}_{5}$  of 
$<C^{-1}A^{-1}\phi^{0}\phi^{0}>$ S-matrix, we then obtain the following elements of string amplitude

\beqa
&& -2\xi_{3}.\xi_{2}k_{2b}k_{3c}\xi_{1a}\pi^2\mu_p\frac{16}{(p-2)!}\eps^{a_{0}\cdots a_{p-3}cba}H_{a_{0}\cdots a_{p-3}} \nonumber\\&&\times
\bigg(\sum_{n=-1}^{\infty}b_n\bigg(\frac{1}{u}(t+s)^{n+1}\bigg)
            +\sum_{p,n,m=0}^{\infty}e_{p,n,m}u^{p}(st)^{n}(s+t)^m\bigg)\label {newcop}\eeqa
            
            where we have already produced all the u-channel poles, now to obtain the new couplings , we need to focus  on  the second term in \reef{newcop} as 
\beqa
&& -2\xi_{3}.\xi_{2}k_{2b}k_{3c}\xi_{1a}\pi^2\mu_p\frac{16}{(p-2)!}\eps^{a_{0}\cdots a_{p-3}cba}H_{a_{0}\cdots a_{p-3}} 
\sum_{p,n,m=0}^{\infty}e_{p,n,m}u^{p}(st)^{n}(s+t)^m\label {newcop3}\eeqa

where \reef{newcop3} satisfies the Ward identity associated to the gauge field,  which means that by replacing $\xi_{1a}$ to $k_{1a}$, apply the momentum conservation and taking the following identity for RR 
 \beqa
 p^a\eps^{a_{0}\cdots a_{p-3}cba}=0\nonumber\eeqa
 the amplitude vanishes. Thus we  understand that \reef{newcop3} has to be reconstructed by new coupling and the structure of this new coupling is shown by

\beqa
\int_{\sum_{p+1}}d^{p+1}\sigma  \quad  \Tr(C_{p-3}\wedge F \wedge D\phi^i \wedge D\phi_i)\label{newnew}\eeqa

Note that \reef{newnew} is considered by the fact that it has to cover up the whole world volume space and more crucially it has to be antisymmetric with respect to interchanging the  momenta of both scalar fields. We now apply  $e_{1,0,0}=\frac{\pi^2}{6}$ and
$e_{0,0,1}=\frac{\pi^2}{3}$
to \reef{newcop3} to be able to start constructing  new couplings at order of $\alpha'^3$.

Indeed if we replace $e_{1,0,0}=\frac{\pi^2}{6}$ to \reef{newcop3} and consider the above remarks, then one can show that,  this term of S-matrix can be generated by the following new coupling as follows

\beqa
S_{12}&=&\frac{(2\pi\alpha')^3\mu_p\pi}{12}\int d^{p+1}\sigma {1\over (p-3)!}(\veps^v)^{a_0\cdots a_{p}}
\left(\frac{\alpha'}{2}\right)
\nonumber\\&&\times C^{(p-3)}_{a_0\cdots a_{p-4}}\Tr\bigg( F_{a_{p-3}a_{p-2}} (D^aD_a)  \bigg[D_{a_{p-1}}\phi^i D_{a_{p}}\phi_i\bigg]\bigg)
\labell{hderv5721}
\eeqa

Notice that, if we do the same for  $e_{0,0,1}=\frac{\pi^2}{3}$, namely if we replace $e_{0,0,1}=\frac{\pi^2}{3}$ into \reef{newcop3} then one gets to know that, this particular term of S-matrix can be obtained by the following new coupling

\beqa
S_{13}&=&\frac{(2\pi\alpha')^3\mu_p\pi}{6}\int d^{p+1}\sigma
 \left(\alpha'\right)
\Tr\bigg(C_{p-3}\wedge D^{b_1} F\wedge  D_{b_1}  \bigg[  D\phi^i \wedge  D\phi_i\bigg]\bigg)
\labell{hderv36}
\eeqa

where these couplings are  of $\alpha'^3$ order.

\vskip.2in

Hence the above couplings \reef{s24r1}, more crucially \reef{hderv5721} and \reef{hderv36} are needed in order to consider the symmetries of the S-matrix with respect to interchanging of the scalar fields.  We can also investigate the closed form of the corrections to all orders in $\alpha'$. So to produce the whole  \reef{newcop3}, one applies the proper higher derivative corrections  to \reef{newnew} so that  the closed form of the string corrections can be found as follows

\beqa
S_{14}&=&\frac{\lambda^3\mu_p}{2\pi}\int d^{p+1}\sigma \sum_{p,n,m=0}^{\infty}
e_{p,n,m}\left(\alpha'\right)^{2n+m}\left(\frac{\alpha'}{2}\right)^{p}
\Tr\bigg(C_{p-3}\wedge D^{b_1}\cdots D^{b_{m}}D^{a_1}\cdots D^{a_{2n}}F\wedge \nonumber\\&& (D^aD_a)^p D_{b_1}\cdots D_{b_{m}} \bigg[ D_{a_1}\cdots D_{a_n}D\phi^i \wedge D_{a_{n+1}}\cdots D_{a_{2n}}D\phi_i\bigg]\bigg)
\labell{hderv367hl}
\eeqa

Note that these new couplings of \reef{hderv5721},\reef{hderv36} and \reef{hderv367hl} can not be derived by the standard effective field theory ways of Taylor, Myers terms nor by pull-back formalism. Indeed not only  the structure of the above new couplings but also their coefficients can  just be explored  by this S-matrix analysis.
\vskip.1in

Note that there is no Ward identity for the amplitudes of scalar fields in the presence of RR, thus  we argue that for two scalars and a  gauge field  in the presence of RR , there is a subtle issue. Indeed to be able to get to the corrected all order contact interactions of higher point functions of string theory amplitudes, one needs to consider both scalar fields in zero picture as we have clarified in detail in the above S-matrix .

\vskip.1in
 It would be nice to generalize this conjecture to even number of scalars in the presence of a closed string RR or even it would be nicer to  check it for the non-BPS amplitudes where the first non-trivial amplitude to be carried out  is $<C^{-1}T^{0}\phi^{-1}\phi^{0}>$ to be compared  with $<C^{-1}T^{-1}\phi^{0}\phi^{0}>$ S-matrix. It would be even more significant if we could carry out these S-matrices  on asymmetric picture of RR   ($<C^{-2}T^{0}\phi^{0}\phi^{0}>$). It is more crucial to actually deal with  the higher point mixed RR- scalar field massless strings to actually generalize the rules and symmetries of string theory amplitudes. We hope to answer these higher point functions of  string amplitudes   and the other issues in  future works.  Although an interesting proposal for picture changing operator has been appeared in \cite{Sen:2015hia}, however,  we find it complicated to be applied  to the real string amplitudes, nevertheless , it would be great to find the deep connections behind those topics as well.

\section {Conclusion}

In this paper, we have  evaluated the  five point world-sheet string theory amplitudes of the mixed RR , scalar and gauge fields, namely we have carried out with entire details the whole  $<C^{-1}\phi ^{0}A^{-1} A^{0}>$, $<C^{-1}\phi ^{-1}A^{0} A^{0}>$, $<C^{-1}A^{0}\phi ^{-1}\phi^{0}>$ and $<C^{-1}A^{-1}\phi ^{0}\phi^{0}>$ S-matrices. 

We have regenerated all  $t,s,u,(t+s+u)$- channel poles in effective field theory.
We also found out  new contact interactions  as well as some new singularities that appear in$<C^{-1}\phi ^{0}A^{-1} A^{0}>$ S-matrix where those new terms were actually the terms that carry momentum of RR in transverse direction and  involved $p.\xi$ terms inside the S-matrix elements.  
These   $p.\xi$ terms  are  needed in the entire form of S-matrix , due to  non zero correlation function of RR field by the first term of scalar field vertex operator in zero picture.  Indeed all $<e^{ip.x(z)} \partial_i x^i(x_1)> $ terms are  non-zero so we have reconstructed the S-matrices such that  by considering all the scalar fields in zero pictures in the presence of RR, we were able  to produce all  $p.\xi$ terms as well as  $ p^{i},p^{j}$ terms ( inside the S-matrices) whose momenta of RR are carried in transverse directions.
By comparing $<C^{-1}\phi ^{0}A^{-1} A^{0}>$ with $<C^{-1}\phi ^{-1}A^{0} A^{0}>$ S-matrix we found a  coupling inside the 
 $<C^{-1}\phi ^{0}A^{-1} A^{0}>$ S-matrix as follows 
 
 \beqa
S_3&=& \frac{i(2\pi\alpha')^3}{2}\mu_p\int d^{p+1}\sigma \quad \Tr(\prt_i C_{(p-3)}\wedge F \wedge F \Phi^i)
\nonumber
\eeqa
 where this coupling can be explained by the effective field theory ways as , the mixed Chern-Simons and Taylor expansion of scalar field was needed. We then generalized its all order higher derivative corrections. We  produced all the new singularities of this S-matrix in section six of this paper as well.
 
 \vskip.1in

 We also compared  $<C^{-1}A^{0}\phi ^{-1}\phi^{0}>$ with $<C^{-1}A^{-1}\phi ^{0}\phi^{0}>$ S-matrix for  all order $\alpha'$  contact interactions  as well as singularities  in both transverse and world volume directions of the S-matrices and that leads to finding out 
 various new couplings in string theory effective actions. First we found the following coupling

 \beqa
S_{11}&=& \frac{i(2\pi\alpha')^3}{2}\mu_p\int d^{p+1}\sigma \quad \Tr(\prt_j\prt_i C_{(p-1)}\wedge F  \Phi^i\Phi^j)
\nonumber
\eeqa 
  and claimed that this coupling can be verified just by $<C^{-1}A^{-1}\phi ^{0}\phi^{0}>$ S-matrix where from field theory we employed the Taylor expansion of scalar fields and then  we generalized its all order corrections.

Basically, we claim that various new contact interactions  appear in the S-matrix by considering both scalar fields in zero picture. Indeed we derived the following new couplings

\beqa
S_{12}&=&\frac{(2\pi\alpha')^3\mu_p\pi}{12}\int d^{p+1}\sigma {1\over (p-3)!}(\veps^v)^{a_0\cdots a_{p}}
\left(\frac{\alpha'}{2}\right)
\nonumber\\&&\times C^{(p-3)}_{a_0\cdots a_{p-4}}\Tr\bigg( F_{a_{p-3}a_{p-2}} (D^aD_a)  \bigg[D_{a_{p-1}}\phi^i D_{a_{p}}\phi_i\bigg]\bigg)
\nonumber
\eeqa

as well as 

\beqa
S_{13}&=&\frac{(2\pi\alpha')^3\mu_p\pi}{6}\int d^{p+1}\sigma
 \left(\alpha'\right)
\Tr\bigg(C_{p-3}\wedge D^{b_1} F\wedge  D_{b_1}  \bigg[  D\phi^i \wedge  D\phi_i\bigg]\bigg)
\nonumber
\eeqa

These couplings  are needed in order to consider the symmetries of the S-matrix with respect to interchanging of the scalar fields and their all order   $\alpha'$ corrections generalized  in  \reef{hderv367hl}.

\vskip.1in

Note that these two above couplings can not be derived by the standard effective field theory ways of Taylor, Myers terms nor by pull-back formalism. Indeed not only  the structures of the above new couplings but also their coefficients can  just be explored  by  $<C^{-1}A^{-1}\phi ^{0}\phi^{0}>$
S-matrix analysis and not by any other tools. 

 \vskip.1in
Note that there is no Ward identity for the amplitudes of scalar fields in the presence of RR, thus  we argue that for two scalars and a  gauge field  in the presence of RR , there was a subtle issue. Indeed to be able to get to new couplings as well as the corrected all order contact interactions of higher point functions of string theory amplitudes, one needs to consider both scalar fields in zero picture as we have clarified in detail in this paper. Eventually  we have made use of Myers  terms and the terms  whose RR momenta are embedded in transverse directions, to be able to derive  all the singularity structures of an RR, two scalars and a  gauge field amplitude. 
\section*{Acknowledgments}
I would like to thank A.M.Polyakov, E.Witten, A.Sen, J.Schwarz, C. Hull, A. Tseytlin,  C. Bachas,  N.Arkani-Hamed,  W.Lerche , D.Waldram, L.Alvarez-Gaume, K.S. Narain, P. Vanhove, P. Horava, I. Klebanov, M.Douglas,  W.Siegel,  C. Vafa, H. Verlinde, J. Heckman, T. Damour, N.Lambert, R.Russo, J.Polchinski, M. Kontsevich , A. Sagnotti and S. Ramgoolam for very useful discussions. Part of this paper has been done at Harvard, Simons Center, Caltech, University of California at Berkeley, KITP,   IAS in Princeton and in string theory group at Vienna, Technology University but the completion of this work has been carried out during my visits to ICTP, CERN, specially it has been completed in Institute des Hautes Etudes Scientifiques (IHES) at Bures-sur-Yvette, France and at CERN in Geneva. The author also thanks the very warm hospitality of the theory divisions of CERN , ICTP, Physics and Mathematics  departments at IHES, IAS, Caltech, Simons center at Stony Brook and UC Berkeley.



\begin{thebibliography}{2007}
\bibitem{Polchinski:1995mt}
  J.~Polchinski,``Dirichlet-Branes and Ramond-Ramond Charges,''
  Phys.\ Rev.\ Lett.\  {\bf75}, 4724 (1995)
  [arXiv:hep-th/9510017].
\bibitem{Witten:1995im}
  E.~Witten,``Bound states of strings and p-branes,''
  Nucl.\ Phys.\  B {\bf 460},335 (1996)
  [arXiv:hep-th/9510135].
\bibitem{Polchinski:1996na}
  J.~Polchinski,`` Lectures on D-branes,''
  [arXiv:hep-th/9611050]
  ;
  C.~P.~Bachas,
  ``Lectures on D-branes,''
  [arXiv:hep-th/9806199].

\bibitem{Polchinski:1994fq}
 J.~Polchinski, ``Combinatorics of boundaries in string theory,''
Phys. Rev. {\bf D50},6041 (1994)[arXiv:hep-th/9407031]
;
  J.~Polchinski, S.~Chaudhuri and C.~V.~Johnson,
  ``Notes on D-Branes,''
  [arXiv:hep-th/9602052].
  
\bibitem{Hatefi:2015gwa} 
  E.~Hatefi,
  ``Remarks on the mixed Ramond - Ramond, open string scattering amplitudes of BPS, non-BPS and brane- anti-brane,''
  Eur.\ Phys.\ J.\ C {\bf 75}, no. 11, 517 (2015)
    [arXiv:1502.06536 [hep-th]].

\bibitem{Myers:1999ps}
  R.~C.~Myers,``Dielectric-branes,''
  JHEP {\bf 9912}, 022 (1999)
  [arXiv:hep-th/9910053].
\bibitem{Hatefi:2012zh}
  E.~Hatefi,
 ``Shedding light on new Wess-Zumino couplings with their corrections to all orders in alpha-prime,''
  JHEP {\bf 1304}, 070 (2013)
  [arXiv:1211.2413 [hep-th]].

\bibitem{Hatefi:2012bp}
  E.~Hatefi, A.~J.~Nurmagambetov and I.~Y.~Park,
  ``ADM reduction of IIB on $\mathcal{H}^{p,q}$ to dS braneworld,''
  JHEP {\bf 1304}, 170 (2013)
  [arXiv:1210.3825 [hep-th]].


\bibitem{Hatefi:2012sy}
  E.~Hatefi, A.~J.~Nurmagambetov and I.~Y.~Park,
  ``$N^3$ entropy of $M5$ branes from dielectric effect,''
  Nucl.\ Phys.\ B {\bf 866}, 58 (2013)
  [arXiv:1204.2711 [hep-th]]



\bibitem{Howe:2006rv}
  P.~S.~Howe, U.~Lindstrom and L.~Wulff,
   `On the covariance of the Dirac-Born-Infeld-Myers action,''
  JHEP {\bf 0702}, 070 (2007)
  [hep-th/0607156].
\bibitem{Leigh:1989jq}
  R.~G.~Leigh,
  ``Dirac-Born-Infeld Action from Dirichlet Sigma Model,''
  Mod.\ Phys.\ Lett.\ A {\bf 4}, 2767 (1989).

\bibitem{Cederwall:1996pv}
  M.~Cederwall, A.~von Gussich, B.~E.~W.~Nilsson and A.~Westerberg,
  Nucl.\ Phys.\ B {\bf 490}, 163 (1997)
  [hep-th/9610148]
  ;
  M.~Aganagic, C.~Popescu and J.~H.~Schwarz,
  Phys.\ Lett.\ B {\bf 393}, 311 (1997)
  [hep-th/9610249]
;
  M.~Aganagic, C.~Popescu and J.~H.~Schwarz,
  Nucl.\ Phys.\ B {\bf 495}, 99 (1997)
  [hep-th/9612080]
  ;
  M.~Cederwall, A.~von Gussich, B.~E.~W.~Nilsson, P.~Sundell and A.~Westerberg,
  Nucl.\ Phys.\ B {\bf 490}, 179 (1997)
  [hep-th/9611159]
  ;
  E.~Bergshoeff and P.~K.~Townsend,
  ``Super D-branes,''
  Nucl.\ Phys.\ B {\bf 490} (1997) 145
  [hep-th/9611173].
\bibitem{Hatefi:2010ik}
  E.~Hatefi,
  ``On effective actions of BPS branes and their higher derivative
  corrections,''
  JHEP {\bf 1005}, 080 (2010)
  [arXiv:1003.0314 [hep-th]].

\bibitem{Hatefi:2012wj}
  E.~Hatefi,
  ``On higher derivative corrections to Wess-Zumino and Tachyonic actions in type II super string theory,''
  Phys.\ Rev.\ D {\bf 86}, 046003 (2012)
  [arXiv:1203.1329 [hep-th]].


\bibitem{Hashimoto:1996bf}
A.~Hashimoto and I.~R.~Klebanov,
``Scattering of strings from D-branes,''
Nucl.\ Phys.\ Proc.\ Suppl.\  {\bf 55B}, 118 (1997)
[arXiv:hep-th/9611214]
 ;
 A. Hashimoto and I. R. Klebanov ,
 ``Decay of excited D-branes,''
 Phys. Lett.~{\bf B381}, 437 (1996) [arXiv:hep-th/9604065]
 ;
S.S.\ Gubser, A.\ Hashimoto, I.R.\ Klebanov, and J.M.\ Maldacena,
;
C.\ Bachas, ``D-Brane Dynamics,'' Phys. Lett.~ {\bf B374}, 37 (1996)[arXiv:hep-th/9511043]
;
  J.~Polchinski,``String duality: A colloquium,''
  Rev.\ Mod.\ Phys.\  {\bf 68}, 1245 (1996) [arXiv:hep-th/9607050];
  W.~Taylor,``Lectures on D-branes, gauge theory and M(atrices),''
  [arXiv:hep-th/9801182];
  N.~D.~Lambert, H.~Liu and J.~M.~Maldacena,
  JHEP {\bf 0703}, 014 (2007)
  [hep-th/0303139].
  C.~Vafa,``Lectures on strings and dualities,''
  [arXiv:hep-th/9702201]
;
  E.~Hatefi, A.~J.~Nurmagambetov and I.~Y.~Park,
  Int.\ J.\ Mod.\ Phys.\ A {\bf 27}, 1250182 (2012)
  [arXiv:1204.6303 [hep-th]].
  M.~Billo, M.~Frau, F.~Lonegro and A.~Lerda,
  JHEP {\bf 0505},047 (2005)
  [arXiv:hep-th/0502084]
  ;
  M.~Billo, P.~Di Vecchia, M.~Frau, A.~Lerda, I.~Pesando, R.~Russo and S.~Sciuto,
  Nucl.\ Phys.\  B {\bf 526},199 (1998)
  [arXiv:hep-th/9802088];
  S.~de Alwis, R.~Gupta, E.~Hatefi and F.~Quevedo,
  JHEP {\bf 1311} (2013) 179
  [arXiv:1308.1222 [hep-th], arXiv:1308.1222].
;
I.~Y.~Park,``Open string engineering of D-brane geometry,''
  JHEP {\bf 0808}, 026 (2008)
  [arXiv:0806.3330[hep-th]].

\bibitem{Hatefi:2012ve}
  E.~Hatefi and I.~Y.~Park,
  ``More on closed string induced higher derivative interactions on D-branes,''
  Phys.\ Rev.\ D {\bf 85}, 125039 (2012)
  [arXiv:1203.5553 [hep-th]].


\bibitem{Park:2008sg}
  I.~Y.~Park,
  ``One loop scattering on D-branes,''
  Eur.\ Phys.\ J.\ C {\bf 62}, 783 (2009)
  [arXiv:0801.0218 [hep-th]].

\bibitem{Hatefi:2015okf} 
  E.~Hatefi,
  ``On RR Couplings and Bulk Singularity Structures of Non-BPS Branes,''
  arXiv:1511.04971 [hep-th].

\bibitem{Hatefi:2012rx}
  E.~Hatefi and I.~Y.~Park,
  ``Universality in all-order $\alpha'$ corrections to BPS/non-BPS brane world volume theories,''
  Nucl.\ Phys.\ B {\bf 864}, 640 (2012)
  [arXiv:1205.5079 [hep-th]].

\bibitem{Kennedy:1999nn}
  C.~Kennedy and A.~Wilkins,``Ramond-Ramond couplings on brane-antibrane systems,''
  Phys.\ Lett.\  B {\bf 464}, 206 (1999)
  [arXiv:hep-th/9905195];
  E.~Hatefi,``SYM, Chern-Simons, Wess-Zumino Couplings and their higher derivative corrections in IIA Superstring theory,''
  Eur.\ Phys.\ J.\ C {\bf 74}, 2949 (2014)
  [arXiv:1403.1238 [hep-th]];.
  E.~Hatefi,``More on Ramond-Ramond, SYM, WZ couplings and their corrections in IIA,''
  Eur.\ Phys.\ J.\ C {\bf 74}, no. 10, 3116 (2014)
  [arXiv:1403.7167 [hep-th]].

\bibitem{Chandia:2003sh}
  O.~Chandia and R.~Medina,``4-point effective actions in open and closed superstring theory,''
  JHEP {\bf 0311},003 (2003)
  [arXiv:hep-th/0310015]
  ;
  L.~A.~Barreiro and R.~Medina,``5-field terms in the open superstring effective action,''
  JHEP {\bf 0503},055 (2005)
  [arXiv:hep-th/0503182];
  E.~Hatefi,
  Nucl.\ Phys.\ B {\bf 880}, 1 (2014)
  [arXiv:1302.5024 [hep-th]];
  E.~Hatefi,
  JHEP {\bf 1307}, 002 (2013)
  [arXiv:1304.3711 [hep-th]].
  R.~Medina, F.~T.~Brandt and F.~R.~Machado,``The open superstring 5-point amplitude revisited,''
  JHEP {\bf 0207},071 (2002)
  [arXiv:hep-th/0208121].

\bibitem{Hatefi:2013yxa}
  E.~Hatefi,
  ``Selection Rules and RR Couplings on Non-BPS Branes,''
  JHEP {\bf 1311}, 204 (2013)
  [arXiv:1307.3520];
  E.~Hatefi,
  ``On D-brane anti D-brane effective actions and their corrections to all orders in alpha-prime,''
  JCAP {\bf 1309} (2013) 011
  [arXiv:1211.5538 [hep-th]];
  M.~R.~Garousi and E.~Hatefi,
  Nucl.\ Phys.\  B {\bf 800}, 502 (2008)
  [arXiv:0710.5875 [hep-th]];
  M.~R.~Garousi and E.~Hatefi,
  JHEP {\bf 0903}, 08 (2009)
  [arXiv:0812.4216 [hep-th]].

\bibitem{Bianchi:1991eu}
  M.~Bianchi, G.~Pradisi and A.~Sagnotti,
  ``Toroidal compactification and symmetry breaking in open string theories,''
  Nucl.\ Phys.\ B {\bf 376}, 365 (1992).

\bibitem{Liu:2001qa}
  H.~Liu and J.~Michelson,
  Nucl.\ Phys.\  B {\bf 614}, 330 (2001)
  [arXiv:hep-th/0107172];
  ;
  E.~Hatefi,
  Eur.\ Phys.\ J.\ C {\bf 74}, no. 8, 3003 (2014)
  [arXiv:1310.8308 [hep-th]];
\bibitem{Hatefi:2015jpa} 
  E.~Hatefi,
  ``On Singularity Structure, New RR-String Couplings in Asymmetric Picture and Their All Order $\alpha'$ Corrections,''
  arXiv:1507.02641 [hep-th].
  
\bibitem{Fotopoulos:2001pt}
  A.~Fotopoulos,``On (alpha')**2 corrections to the D-brane action for non-geodesic
  world-volume embeddings,''
  JHEP {\bf 0109}, 005 (2001)
  [arXiv:hep-th/0104146].
\bibitem{Sen:2015hia} 
  A.~Sen and E.~Witten,
  ``Filling the gaps with PCOs,''
  JHEP {\bf 1509}, 004 (2015)
  [arXiv:1504.00609 [hep-th]].
\end{thebibliography}
\end{document}